\documentclass[final,5p,times,square,sort&compress]{elsarticle}

\usepackage{amssymb}
\usepackage{amsmath}

\usepackage{caption}
\usepackage{graphicx}%
\usepackage{multirow}%
\usepackage{amsmath,amssymb,amsfonts}%
\usepackage{amsthm}%
\usepackage{mathrsfs}%
\usepackage[title]{appendix}%
\usepackage{xcolor}%
\usepackage{textcomp}%
\usepackage{manyfoot}
\usepackage{array}
\usepackage{balance}
\usepackage{booktabs}%
\usepackage{algorithm}%
\usepackage{algpseudocode}%
\usepackage{listings}%
\usepackage{geometry}
\usepackage{multirow}
\usepackage{subcaption}
\usepackage{todonotes}
\usepackage{enumitem}
\usepackage{comment}
\usepackage{xcolor}
\usepackage{url}
\algnewcommand\INPUT{\item[\textbf{Input:}]}
\algnewcommand\OUTPUT{\item[\textbf{Output:}]}

\makeatletter
\def\ps@pprintTitle{%
 \let\@oddhead\@empty
 \let\@evenhead\@empty
 \let\@oddfoot\@empty
 \let\@evenfoot\@oddfoot
}
\makeatother

\usepackage{adjustbox}
\usepackage{pifont}

\begin{document}

\begin{frontmatter}



\title{Hierarchical Graph Neural Network \textcolor{black}{for} Compressed Speech Steganalysis} 


\author[label1]{Mustapha Hemis} 
\ead{mhemis@usthb.dz}
\author[label2]{Hamza Kheddar}
\ead{kheddar.hamza@univ-medea.dz}
\author[label3]{Mohamed Chahine Ghanem *~}
\ead{mohamed.chahine.ghanem@liverpool.ac.uk}
\author[label1]{Bachir Boudraa}
\ead{bboudraa@usthb.dz}

\affiliation[label1]{organization={LCPTS Laboratory, University of Sciences and Technology Houari Boumediene (USTHB)},
            city={El-Alia, Bab-Ezzouar},
            postcode={16111}, 
            state={Algiers},
            country={Algeria}}

\affiliation[label2]{organization={LSEA Laboratory, Department of Electrical Engineering, University of Medea},
            postcode={26000},
            state={Medea},
            country={Algeria}}

\affiliation[label3]{organization={Cybersecurity Institute, School of Computer Science and Informatics, University of Liverpool},
            postcode={L69 3BX},
            state={Liverpool},
            country={UK}}

\affiliation[label4]{organization={Cyber Security Research Centre, School of Computing and Digital Media, London Metropolitan University},
            postcode={L69 3BX}, 
            state={London},
            country={UK}}

\begin{abstract}
Steganalysis methods based on deep learning (DL) often struggle with computational complexity and challenges in generalizing across different datasets. {In the specific case of voice-over-IP (VoIP) speech streams, detection is particularly challenging because the low bit-rate encoding creates complex, relational dependencies between speech frames. Conventional DL models, which treat data as simple sequences or grids, often fail to capture these complex inter-frame dependencies effectively. To address this gap,} this paper presents the first application of a graph neural network (GNN), specifically the GraphSAGE architecture, for steganalysis of compressed VoIP speech streams.
The method involves straightforward graph construction from VoIP streams and employs GraphSAGE to capture hierarchical steganalysis information, including both fine-grained details and high-level patterns, thereby achieving high detection accuracy. Experimental results demonstrate that the developed approach performs well in uncovering quantization index modulation (QIM)-based steganographic patterns in VoIP signals. It achieves detection accuracy exceeding 98\% even for short 0.5-second samples, and 95.17\% accuracy under challenging conditions with low embedding rates, representing an improvement of 2.8\% over the best-performing state-of-the-art methods. Furthermore, the model exhibits superior efficiency, with an average detection time as low as 0.016 seconds for 0.5-second samples—an improvement of 0.003 seconds. This makes it efficient for online steganalysis tasks, providing a superior balance between detection accuracy and efficiency under the constraint of short samples with low embedding rates.
\end{abstract}


\end{frontmatter}



\section{Introduction}
\label{section1}

Data hiding has garnered significant attention from researchers in recent years, spanning various disciplines. On the one hand, it serves to verify integrity and ensure information remains unaltered during transmission, as seen in watermarking \cite{hemis2018adjustable, rebahi2023image}. On the other hand, it can be utilized to conceal confidential information through unconventional means, exemplified by steganography \cite{kheddar2022high,abdelmged2016improving}. Steganography and steganalysis are two complementary aspects of covert communication. While steganography involves concealing secret information within apparently benign carriers such as images \cite{megias2022subsequent}, text \cite{khosravi2019new}, and speech \cite{kheddar2022high,kheddar2018fourier,kheddar2022speech}, steganalysis is focused on detecting and unveiling such concealed communication \cite{kheddar2024deep}. Over the years, the proliferation of Internet services has facilitated the transmission of multimedia data over digital networks. Among these services, voice over internet protocol (VoIP) has emerged as a prominent means of enabling real-time voice communication over the Internet. This increased reliance on VoIP, driven by the success of platforms like Skype, WhatsApp, and Zoom, has made it an attractive choice for concealing and transmitting hidden information \cite{tian2014improving}. The ubiquity and high-volume nature of VoIP traffic make it an appealing vector for steganographic purposes, presenting both opportunities for clandestine communication and challenges for cybersecurity professionals.

In VoIP communications, speech is typically compressed using codecs, such as G.722, G.723, and G.729, to reduce bandwidth requirements. This compression process, which involves the quantization of speech parameters, inadvertently creates potential vulnerabilities that can be exploited for steganographic purposes. Malicious actors, including cybercriminals and extremist groups, may exploit these vulnerabilities by manipulating VoIP software to facilitate covert communications. This scenario presents a significant challenge to communication monitoring and network security. Steganography for VoIP can occur both in network protocol fields and payload data \cite{wu2021steganography, kheddar2019pitch}. However, while the former approach may provide some degree of concealment, it is the latter that takes precedence. This distinction arises from the fact that the practice of concealing information within network protocol fields, which encompass many layers of the open systems interconnection (OSI) model, ultimately provides less robust concealment capabilities. These protocol fields are typically public, and their data tends to remain fixed under most ordinary circumstances, rendering them less secure for covert communication \cite{yang2019hierarchical}. On the contrary, the method of embedding data within payload data, characterized by its dynamic nature and temporal variations, offers a substantially higher degree of concealment. This dynamism makes it considerably more challenging to detect covert practices within VoIP streams, reinforcing its position as the preferred choice for steganographic endeavors \textcolor{black}{\cite{yang2019hierarchical, yang2022real}}. Additionally, as steganography methods become more sophisticated, particularly with the advent of AI-generated content, advanced steganalysis techniques are necessary to keep pace \textcolor{black}{\cite{li2022general}}.

As a result, there is a compelling need to develop effective steganalysis methods tailored specifically for VoIP steganography. These methods can address several critical concerns, including: (i) For national security agencies, detecting and preventing the covert exchange of sensitive information via VoIP is crucial to national security; (ii) businesses depend on steganalysis to safeguard against intellectual property theft and unauthorized data exfiltration through VoIP channels, protecting their proprietary information and maintaining a competitive advantage; (iii) many industries are required to monitor and secure their communications to comply with regulatory standards, making VoIP steganalysis essential for ensuring compliance with legal and industry-specific requirements.

An effective steganalysis in the context of VoIP streams should meet two crucial requirements \cite{lin2018rnn, ghane2023novel}. Firstly, it should operate in real-time, ensuring that the time required for detection is minimized due to the need for swift action against potential malicious activities. Secondly, the steganalysis method must be capable of detecting short samples of VoIP streams, as covert information may be hidden within brief segments of the communication. Adding to these two requirements, it should be sensitive enough to detect low embedding rates, as some covert data may be hidden with minimal changes to the host signal. \textcolor{black}{Meeting these stringent requirements represents a significant and persistent challenge in developing practical and reliable VoIP steganalysis methods, which is essential to ensure their successful integration into real-world applications \cite{lin2018rnn,wu2021steganography}.
In this work, we address the core research problem of designing a steganalysis approach for VoIP that can achieve real-time detection performance, maintain high accuracy on short speech segments, and remain effective under low embedding rates.}

\subsection{Motivation}

The steganalysis process generally involves a two-step process: first, identifying distinguishing features from the carrier signal, and then classifying whether it contains steganographic content. Feature extraction involves identifying and selecting key pieces of data from the carrier, such as patterns or anomalies, that can indicate the presence of steganography. Traditional methods rely on handcrafted statistical features, designed based on expert knowledge, to detect significant changes introduced by steganographic embedding. However, these methods face limitations in the context of VoIP due to the minimal additional distortion introduced by steganography in compressed speech, making it challenging to extract suitable features for steganalysis.

With the rapid advancement of deep learning (DL), recent methods have leveraged deep models to automatically learn discriminative features. Various neural network architectures, primarily leveraging convolutional neural network (CNN)- and recurrent neural network (RNN)-based designs \cite{qiu2022steganalysis,lin2018rnn, yang2022real,wu2023mfpd}, have been explored in VoIP steganalysis. These models are specifically designed to capture distinct codeword correlation features between cover and stego VoIP streams, achieving state-of-the-art (SOTA) detection outcomes. \textcolor{black}{While effective in capturing sequential (RNNs) or local spatial (CNNs) patterns, these traditional deep learning architectures face inherent limitations when dealing with the relational structure of compressed VoIP data affected by steganography. QIM steganography, in particular, subtly alters the dependencies and correlations between line spectral frequency (LSF) codewords across frames and within their neighborhoods. CNNs, designed for grid-like data, struggle to explicitly model these non-Euclidean, inter-frame relationships. RNNs, while adept at sequences, might not efficiently capture the graph-like dependencies that extend beyond simple linear order or local windows, especially when the crucial information lies in the way codewords relate to each other rather than just their individual values or linear sequence.}
	
Recently, significant attention has been directed toward adapting DL methods for graph-structured data, giving rise to the the emergence of graph neural networks (GNNs) as a prominent topic \cite{ wu2020comprehensive}. \textcolor{black}{GNNs are uniquely suited for modeling complex relationships within non-Euclidean data by explicitly learning from graph structures, where nodes represent entities (e.g., LSF codeword frames) and edges represent their relationships (e.g., temporal dependencies). This intrinsic capability allows GNNs to capture both fine-grained local dependencies and high-level global patterns by aggregating information from connected neighborhoods.}
GNNs have demonstrated efficiency in representing and analyzing graph data across various computer vision fields, including action recognition \cite{geng2024hierarchical}, object tracking \cite{zhan2024gnn}, and natural language processing, such as text classification \cite{peng2024novel}. However, there have been limited efforts in applying GNNs to steganalysis, primarily in the domains of images, as in \cite{liu2022graph, liu2023jpeg}, and text, as in \cite{wu2021linguistic, fu2022hga}. \textcolor{black}{This highlights a significant gap where GNNs' relational modeling power could be particularly advantageous for VoIP steganalysis.}

\subsection{Contribution and paper structure}

DL-based steganalysis commonly encounter trade-offs between performance and processing complexity. This study introduces the application of a GraphSAGE GNN network to design an effective approach that meets the essential requirements of VoIP steganalysis systems. The key contributions of this study can be outlined as follows:

\setlength{\leftmargini}{10pt}
\begin{itemize}
\item Propose, according to the literature reviewed, the first application of GNNs in the context of VoIP steganalysis, marking a significant advancement in quantization index modulation (QIM)-based steganalysis methodologies.

\item Introduce a simple and efficient approach for constructing graphs from input VoIP streams, resulting in a lightweight and simplified graph structure that reduces computational complexity while preserving the capability for excellent feature extraction, crucial for identifying patterns indicative of steganography.

\item Propose a GNN architecture based on the GraphSAGE model, specifically designed to hierarchically extract steganalysis information. This architecture, for the first time, effectively utilizes LSF codewords from VoIP codecs notably G.723 and G.729 to capture both fine-grained details and high-level patterns, resulting in high detection accuracy.

\item Prove, through experiments, that the suggested GNN-based speech steganalysis technique achieves competitive detection accuracy in the challenging scenario. \textcolor{black}{Specifically, it attains detection accuracy exceeding 98\% even for short 0.5-second samples, and 95.17\% accuracy under challenging conditions with low embedding rates, representing an improvement of 2.8\% over the best-performing state-of-the-art methods. Furthermore, the model exhibits superior efficiency, with an average detection time as low as 0.016 seconds for 0.5-second samples—an improvement of 0.003 seconds. These results highlight the model’s suitability for real-time deployment, offering a robust trade-off between detection accuracy and computational efficiency.}

\color{black}

\item  Demonstrate  practical implications by enabling real-time detection of covert communications in VoIP streams for real-world scenarios. This includes enhanced cybersecurity and network monitoring, support for law enforcement in identifying hidden transmissions, and applications in digital rights management and IoT security, thanks to the method’s high detection accuracy and efficiency.
\color{black}
\end{itemize}

The remainder of this paper is structured as follows: Section \ref{section2} delves into related works. Section \ref{section3} offers an overview of the background. Section \ref{section4} introduces the proposed proposed GNN-based steganalysis framework. Section \ref{section5} presents experimental results and offers an in-depth examination of the results. Section \ref{section6} outlines the key limitations and shortcomings of the approach, suggests possible enhancements, and discusses the applicability of the proposed scheme in several real-world scenarios. Finally, section \ref{section7} presents conclusive remarks and outlines future directions.

\section{Related work}\label{section2}
	
Vocoders, such as G.723 and G.729, are widely applied in VoIP communication with the aim of lowering decoding errors via the analysis-by-synthesis (AbS) approach, maintaining excellent speech quality while achieving high compression ratios. In contrast to conventional steganographic methods typically used for images or text, low bit-rate (LBR) VoIP streams, characterized by the AbS linear predictive coding technique, pose a unique challenge due to their minimal redundancy within the encoded speech. To address this, researchers have explored two distinct categories of data hiding techniques, each linked to different stages in the encoding process. The first category involves altering specific elements in the compressed speech stream, often utilizing methods like least significant bit (LSB) replacement, such as in  \cite{kheddar2019pitch}. However, this is complicated due to the limited data-hiding space in LBR speech. The second category focuses on embedding covert information during encoding, by altering encoding features, notably the linear predictive coding (LPC) \cite{laskar2024enhancing}, fixed codebook (FCB) \cite{liu2016neighbor}, and adaptive codebook (ACB) \cite{shufan2015steganography}.

In the second category, QIM-based steganography is most commonly employed, especially within LPC domains. QIM steganography, originally developed by Chen et al. \cite{chen2001quantization}, is widely used in literature. It conceals steganograms by altering the quantization vector within the speech code, introducing minimal  unperceptible distortion and offering high data hiding capabilities. Thus, when it comes to embedding information in VoIP streams, QIM emerges as a suitable scheme that poses a significant challenge for detection. 

On the one hand, research into QIM steganography detection methods has primarily concentrated on image carriers \cite{malik2011nonparametric}. Subsequently, a limited number of research studies have been proposed in the field of steganalysis, particularly focusing on QIM-based steganography in VoIP, such as in \cite{yang2019hierarchical, sun2021audio}. Due to the distinct characteristics of VoIP streams, the detection methods proposed for use in images cannot be directly applied within the VoIP context. Moreover, classical audio steganalysis methods based on statistical feature extraction in the uncompressed domain are inapplicable to VoIP. This is due to the minimal additional distortion introduced by VoIP QIM steganography in decoded speech signals, making it challenging to extract suitable features for steganalysis \cite{yang2019steganalysis}. To address this issue, some research in the VoIP field has proposed the use of handcrafted statistical features to detect significant changes in codewords used in LBR speech codec streams caused by QIM steganography.  
For instance, Li et al. \cite{li2012steganalysis} examined the impact of QIM steganography on G.729, where it altered the quantization index of the LPC filter. They employed a statistical model to capture codeword distribution characteristics and, in conjunction with a support vector machine (SVM), formulated an efficient system for steganographic detection. In a related study, Li et al. \cite{li2012detection} developed the index distribution characteristics (IDC) steganalysis method, employing codeword distribution histograms and first-order Markov chain-derived state transition probabilities as correlation features. These features, integrated with SVM, were utilized to classify steganographic samples with a focus on inter-frame transition probabilities. Expanding upon this concept, the studies by Li et al. \cite{li2017steganalysis} and Wu et al. \cite{wu2020steganalysis} aim to detect steganograms embedded using QIM techniques. Both works employ principal component analysis (PCA) for reducing feature dimensionality and SVM for classification.  Notably, Li et al. \cite{li2017steganalysis} focus on state transition probabilities of intra-frame codewords as features, while Wu et al. \cite{wu2020steganalysis} suggested extracting features from the original speech signal by analyzing its distribution patterns and transmission likelihoods. Meanwhile, Yang et al. \cite{yang2018steganalysis} employed temporal dependencies between codewords across frames to model a Codeword Bayesian Network (CBN), where the parameter learning process was guided by a Dirichlet prior.  Yang et al. \cite{yang2019fast} proposed a fast steganalysis which involved mapping vector quantization codewords to a semantic space, followed by feature extraction using a hidden layer input into a softmax classifier. Such methods require meticulous feature design to align with the characteristics of speech data and specific information hiding techniques. Consequently, they struggle to effectively address generative steganography that produces high-quality stego speech.   
On the other hand, the proposed GraphSAGE-based algorithm addresses these challenges by utilizing relational data through GNNs and efficiently managing the minimal distortion and short sample lengths typical of VoIP streams.

Besides, recent advancements in DL have led to improved robustness and effectiveness in steganalysis methods, which can be broadly categorized into two main categories: general steganalysis approaches and specific steganalysis approaches. In the first category, several methods have been introduced to detect a range of steganographic techniques. For example, Hu et al. \cite{hu2021detection} addressed heterogeneous parallel steganography (HPS) by introducing the steganalysis feature fusion network (SFFN). The architecture of SFFN comprises three core modules: a network for feature extraction, a component for feature integration, and a classification unit,  effectively extracting and combining steganalysis features from HPS methods for dependable predictions. Building upon this, Wang et al. \cite{wang2021fast} proposed a fast and efficient steganalysis method which leveraged attention mechanisms. Their suggested method adeptly extracts key characteristics related to steganography exceptions and fuses these features targeting multiple steganography techniques  utilized in HPS. Li et al. \cite{li2021detection} introduced a unified detection framework, termed CBCA, which leverages codeword embedding along with  \textcolor{black}{bidirectional long short-term memory (Bi-LSTM)} and CNN-based attention mechanisms. This approach is capable of identifying three distinct steganographic techniques concurrently. Wang et al. \cite{wang2022steganalysis} proposed a steganalysis method aimed at identifying various steganographic schemes within compressed speech signals. Their framework introduces a codeword-distributed embedding component designed to generate compact representations from compressed codewords. To capture dependencies across different contexts, it employs two correlation extraction modules: a global-guided unit that incorporates Bi-LSTM and multi-head self-attention, and a local-guided unit built using the convolutional block Attention module (CBAM), enabling the modeling of both global and local correlations before and after data hiding.  Detection is accomplished through fully connected layers, ensuring accurate identification. Li et al. \cite{li2022general} introduced a generic frame-level steganalysis approach for LBR steganography, utilizing dual-domain feature representation and a Transformer-based model named Stegaformer.  The proposed method consists of two modules, dual-domain representation and Stegaformer. Li et al. \cite{li2023sanet} introduced SANet, an independent steganalysis network for speech encoding and steganography. This approach  unifies  compressed speech outputs from various codecs by converting them into a common uncompressed-domain format, introducing an intermediate representation. A Bi-LSTM neural network is developed to extract steganography-sensitive characteristics through collaborative correlation features. This approach achieves SOTA detection performance for various steganography algorithms across different speech encoders. \textcolor{black}{Recently, Wang et al. \cite{wang2025swan} proposed E-SWAN, a deep learning-based sliding window analysis network that combines LSTM and convolutional modules for real-time VoIP speech steganalysis. The model demonstrates competitive performance in detecting two steganography methods. Similarly, Lin et al. \cite{li2025compressed} introduced a steganalysis approach that integrates Bi-LSTM, 3D convolution  and attention mechanisms to detect two QIM-based steganography variants in VoIP streams.}
Although these algorithms offer some capability for broad steganalysis applications, their performance often falters when confronted with specific steganographic techniques, particularly QIM. In such specialized scenarios, these general methods typically exhibit lower accuracy, highlighting the need for more targeted approaches.

Consequently, several researchers prefer to exploit the power of DL to design specialized steganalysis algorithms, particularly geared toward QIM-based steganography. 
For example, Lin et al. \cite{lin2018rnn} identified four robust codeword correlation patterns and introduced a steganalysis model based on RNN for improved detection accuracy. Nonetheless, their model, RNN-SM, centred on global contextual information, utilizing a two-layer LSTM. Yang et al. \cite{yang2019steganalysis} proposed a novel approach to enhance local representations that combined RNN and CNN, resulting in the CNN-LSTM model, which outperformed RNN-SM in the detection of QIM-based steganography. Yang et al. \cite{yang2019hierarchical} incorporated an attention mechanism within a hierarchical convolutional structure, further enhancing steganalysis results. In a subsequent study \cite{yang2020fcem}, the same authors presented a lightweight neural architecture called the fast correlation extract model (FCEM), which integrates positional encoding and multi-head attention to capture correlation features. Qiu et al. \cite{qiu2022steganalysis} developed an efficient steganalysis that incorporates a codeword embedding layer to capture dense representations, employs a bidirectional LSTM layer and incorporates a gated attention module to model contextual feature distributions. Yang et al. \cite{yang2022real} designed a multi-channel convolutional sliding window mechanism to capture inter-frame dependencies between a target frame and its adjacent frames. Recently, the utilization of a transformer encoder \cite{zhang2023tenet,kheddar2025recent,kheddar2025transformers} and federated learning \cite{tian2023fedspy,himeur2023federated} has been implemented for the detection of QIM-based steganography in G.729 speech encoders. These two approaches demonstrated SOTA detection performance. Zhang et al. \cite{zhang2024spm} tackled the challenge of identifying payload locations in QIM-based VoIP steganography by proposing an LSTM-based steganalysis method named SPM. {To provide a structured comparison and highlight the evolution of these key approaches, Table \ref{tab:related_work_summary} presents a summary of the works most relevant to our research, detailing their core techniques and limitations.}

\begin{table*}[h!]
\centering

\caption{{A comparative summary of key SOTA methods in QIM-based VoIP steganalysis, representing the works most closely related to our proposed approach.}}
\label{tab:related_work_summary}
{\fontsize{7.5}{8}\selectfont
\begin{tabular}{p{1.7cm}p{3.2cm}p{4.5cm}p{6.6cm}}
\toprule
\textbf{Ref.} & \textbf{Method / Model} & \textbf{Key contribution} & \textbf{Limitations addressed by our work} \\ 
\midrule

Li et al.  \cite{li2012steganalysis} & Statistical model + SVM & Captures distribution characteristics of LPC quantization indices in G.729 & Requires handcrafted features; limited robustness to low embedding rates  \\ \addlinespace

Li et al. \cite{li2012detection} & Statistical Features + SVM & Used codeword histograms and Markov probabilities to capture inter-frame correlations. & Relies on manually designed features; limited robustness to low embedding rates \\ \addlinespace

Li et al. \cite{li2017steganalysis} & Statistical Features + PCA + SVM & Focused on state transition probabilities of intra-frame codewords. & Requires meticulous feature engineering and is less adaptable than end-to-end models; limited robustness to low embedding rates\\ \addlinespace

Wu et al. \cite{wu2020steganalysis} & Statistical Features + PCA + SVM & Extracted features from the original speech signal's distribution patterns. & Feature design is specific to known signal characteristics; Computationally expensive; weaker for real-time detection. \\  \addlinespace

Yang et al. \cite{yang2018steganalysis} & Codeword bayesian network (CBN) & Modeled temporal dependencies across frames using a probabilistic graphical model. & Relies on a specific statistical model (CBN) that is less flexible than learned representations.\\ \addlinespace

Yang et al.  \cite{yang2019fast} & Semantic mapping + softmax classifier & Projects VQ codewords to semantic space for classification & Lightweight but less accurate for subtle embeddings \\ \addlinespace

Lin et al.  \cite{lin2018rnn} & RNN-SM (Two-layer LSTM) & Captures global contextual correlations & Models data purely sequentially; less effective at capturing complex, non-local dependencies.  \\ \addlinespace

Yang et al.  \cite{yang2019steganalysis} & CNN + LSTM  & Combines local CNN with temporal RNN features & Grid-based,  CNN and sequential models, do not explicitly capture the complex relational structure of VoIP data; Higher complexity \\ \addlinespace

Yang et al.  \cite{yang2019hierarchical} & Hierarchical CNN + Attention & Enhanced detection using multi-level attention & Limited by the inherently local receptive field of CNNs, which struggles to capture complex, long-range dependencies \\ \addlinespace

Yang et al. \cite{yang2020fcem} & FCEM (Lightweight CNN + Attention) & Efficient correlation extraction with positional encoding & Attention works on sequences; weaker in capturing structural relationships \\ \addlinespace

Qiu et al. \cite{qiu2022steganalysis} & Bi-LSTM + gated Attention & Dense contextual feature representation & Sequential model; lacks relational aggregation over graph structures \\ \addlinespace

Zhang et al.  \cite{zhang2024spm} & LSTM-based SPM & Identifies payload locations in VoIP & Task-specific; not generalized for broad steganalysis \\ \addlinespace

Yang et al. \cite{yang2022real} & Multi-channel sliding CNN Window & Captures local inter-frame dependencies between adjacent frames & Sliding window has fixed local receptive field; cannot capture longer-range dependencies \\ \addlinespace

Zhang et al. \cite{zhang2023tenet} & Transformer encoder & Leveraged the power of self-attention from Transformers for steganalysis. & May not inherently model the direct temporal adjacency in speech frames as effectively as a graph structure. \\ \addlinespace

Tian et al. \cite{tian2023fedspy} & Federated learning & Introduced a secure, collaborative learning framework for steganalysis. & Focuses on the training paradigm rather than the core model architecture for feature extraction. \\ \addlinespace

\textbf{Our Work} & \textbf{GNN} & \textbf{First to model VoIP streams as graphs to capture hierarchical relational data.} & \textbf{Explicitly models non-Euclidean, relational dependencies between speech frames, overcoming the limitations of both manual feature engineering and simpler sequential/grid-based models.} \\
\bottomrule
\end{tabular}}
\end{table*}

The mentioned studies underscore the adoption of DL-based methods as the benchmark for real-time VoIP steganalysis, demonstrating superior detection capabilities and efficiency in comparison to handcrafted features. The integration of DL with graph data has led to the widespread adoption of GNNs. These DL-based models, specialized for graph domains, are crafted to capture representations of graph data adeptly, proficiently handling tasks centred on both individual nodes and entire graphs. GNNs capture relationships within graphs by exchanging messages among nodes and maintaining a state that encodes information from any connected neighborhood through edges \cite{zhou2020graph}. Furthermore, GNNs exhibit an enhanced capability to leverage global information, contributing to their compelling performance. While GNNs have been recently employed in various studies for image \cite{liu2022graph, liu2023jpeg} and text steganalysis \cite{wu2021linguistic, fu2022hga, pang2023cats}, there is a noticeable gap in research focusing on steganalysis in the context of VoIP.

This paper aims to investigate the potential of GNNs in crafting steganographic detection approaches that meet the essential requirements of VoIP steganalysis systems. Based on the available literature reviewed by the authors, no existing studies have explored GNN-based VoIP steganalysis. Our work addresses this significant gap by introducing a novel GraphSAGE-based algorithm tailored for VoIP streams, achieving high accuracy and real-time efficiency. Consequently, this approach has the potential for deployment in various real-world applications.

\section{Background}\label{section3}

Developing a novel steganalysis scheme dedicated to VoIP security requires a foundational understanding of how hidden messages, embedded within the bitstream (steganogram), are generated in the coder's output. Furthermore, representing the bitstream in a graph format necessitates prior knowledge of GNNs and an understanding of how speech can be effectively represented in a graphical form. For this purpose, this section summarizes the principles of speech coding, QIM-based steganography, and GNNs.

\subsection{Speech Compression Overview}

In VoIP applications, voice signals are initially compressed using LBR speech encoders at the sender's end before transmission. These encoders are based on the LPC model and operate under the AbS framework, where an LPC filter is used to model and reconstruct speech during both encoding and decoding stages. The LPC filter formulation is given by:

\begin{equation}\label{eq:lpc}
	A(z) = \frac{1}{1 - \sum_{i=1}^{n} a_i z^{-i}},
\end{equation}

Where $A(z)$ is the transfer function of the LPC filter, $z$ is a complex variable that represents the frequency domain, $n$ is the order of the LPC filter, and $a_i$ represents the $i$-th order coefficient of the LPC filter. Since speech exhibits short-term stationarity, it is processed in frames, and LPC coefficients are derived individually for each frame. Due to the high sensitivity of LPC coefficients to quantization noise, they are typically transformed into line spectrum frequencies (LSF), which are easier to quantize to maintain coding stability. LSFs from each frame are quantized using several codebooks based on the minimum mean square error criterion. This quantization aims to select the optimal codeword that reduces the discrepancy between the original and reconstructed speech signals. In particular, G.723 and G.729 implement LSF quantization by choosing codewords \( c_1 \), \( c_2 \), and \( c_3 \) from their corresponding codebooks \( C_1 \), \( C_2 \), and \( C_3 \).

\subsection{QIM-based Steganography}

In the context of QIM steganography, the process of quantizing LSFs can be modified. 
Each original codebook  $C_i$, which is used for vector quantization, is split into two separate parts, $C^1_i$ and $C^2_i$, following the relationship:

\begin{equation}\label{eq:qim}
	\left\{
	\begin{aligned}
		&C_i = C^1_i \cup C^2_i \\
		&C^1_i \cap C^2_i = \emptyset
	\end{aligned}
	\right.
\end{equation}

\begin{figure}
	\centering
	\includegraphics[scale=0.54]{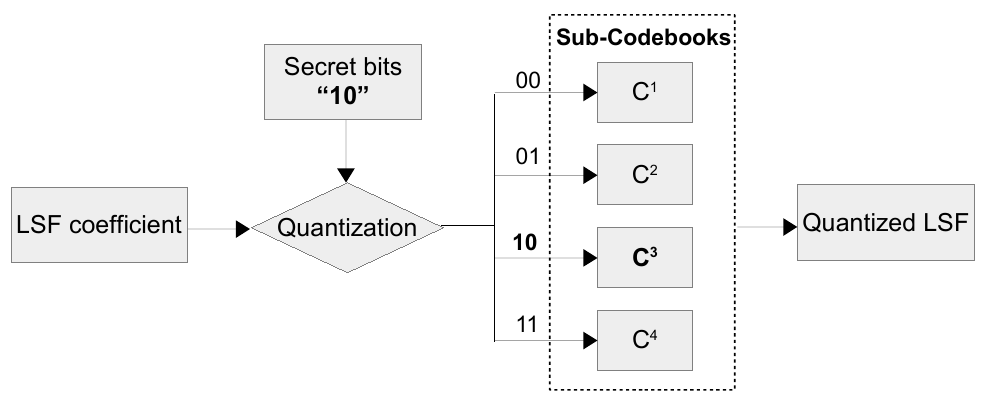}
	\caption{QIM for concealing 2-bit steganograms.}
	\label{fig:qim}
\end{figure}

When embedding the bit "0", the quantizer in the QIM scheme retrieves the closest match from the sub-codebook \( C^1_i \). Conversely, when embedding the bit "1", the QIM method chooses the optimal quantization value from the sub-codebook $C^2_i$. During reception, the receiver can deduce the embedded bits by determining whether the received quantization index corresponds to sub-codebook $C^1_i$ or $C^2_i$. This methodology allows for the seamless integration of secret data within the quantization. In the context of QIM, it is possible to enhance concealment capacity by subdividing the codebook into $2^n$ sub-codebooks, allowing the simultaneous hiding of $n$ bits. The process of embedding and extracting a hidden message with 2 bits length using the QIM technique is depicted in Figure \ref{fig:qim}. This functionality is accomplished by partitioning $C_i$ into sub-codebooks $C^1_i, C^2_i, C^3_i,$ and $C^4_i$, as illustrated in the figure.

The challenge in QIM steganography arises from increased distortion resulting from the use of a reduced codeword set in the quantization procedure. The fundamental issue lies in how to partition the original codebook $C_i$ into multiple sub-codebooks. Researchers have proposed various methods including matrix encoding QIM (ME-QIM) \cite{tian2014improving}, complementary neighbour vertices QIM (CNV-QIM) \cite{xiao2008approach}, and others.

\subsection{GNN Overview}

GNNs are a specialized class of neural networks tailored for learning from graph-structured data. In such data structures, nodes denote entities, while edges capture the relationships or interactions between them. The basic propagation rule in a GNN can be expressed as follows:

\begin{equation}
 h_v^{k} = \mathcal{F} \left( \sum_{u \in N(v)} w_{uv}^{k-1} h_u^{k-1}, h_v^{k-1} \right)   
\end{equation}

\noindent Here, \( h_v^{k} \) and \( h_v^{k-1} \) represent the hidden states of node \( v \) at layers \( k\) and \( k-1\), respectively. The neighborhood of a node \(v\) is symbolized as \(N(v)\), and \( w_{uv}^{k-1} \) represents the weight associated to the edge connecting \( u \) and \( v \) in the \((k{-}1)\)th layer. The function \( \mathcal{F} \) is a non-linear activation function. The update rule essentially aggregates information from neighbouring nodes to refine the current node’s hidden state. 

A widely adopted family of GNN models includes graph convolutional networks (GCNs)~\cite{kipf2016semi},  gated graph neural networks (GGNNs) \cite{li2015gated}, graph sample and aggregation (GraphSAGE) \cite{hamilton2017inductive}, and more. These architectures often involve variations of the basic propagation rule to capture more complex relationships within the graph.

\section{Proposed framework} \label{section4}

\begin{figure*}[ht!]
    \centering    \includegraphics[scale=0.77]{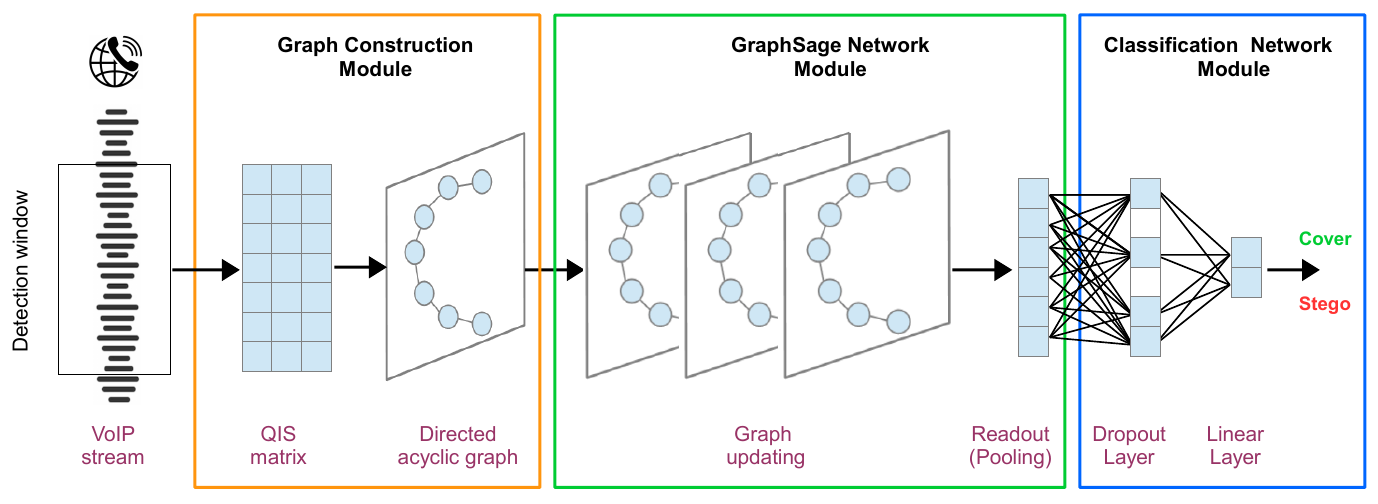}
    \caption{The overall structure of the proposed approach{, which consists of a three-stage pipeline. (1) Graph Construction Module: A QIS matrix is extracted from the input VoIP stream and converted into a directed acyclic graph. (2) GraphSAGE Network Module: The graph is processed via graph updating and pooling to generate a representative feature vector. (3) Classification Network Module: The final vector is passed through dropout and linear layers to classify the stream as 'Cover' or 'Stego'.}}
    \label{Fig:ProposedFramework}
\end{figure*}

\textcolor{black}{Building upon the foundational concepts of speech coding, QIM steganography, and GNN discussed in Section 3, we now present the detailed architecture and components of our proposed steganalysis system. Specifically, this section describes} an efficient method designed to uncover QIM steganography within compressed, LBR VoIP streams. Our approach is specifically tailored to work with speech compressed using the G.729A codec, which is widely used in VoIP communications. Figure~\ref{Fig:ProposedFramework} illustrates the structure of our proposed VoIP steganalysis, while Algorithm \ref{algo1} outlines the steps involved.

As depicted in the figure, our framework consists of a three-stage pipeline. First, the Graph Construction Module takes a raw VoIP stream, extracts its Quantization Index Sequence (QIS) matrix, and converts it into a directed acyclic graph representation. Second, this graph is processed by the GraphSAGE Network Module, which learns a discriminative graph-level feature vector through hierarchical message passing and pooling operations. Finally, the resulting feature vector is fed into the Classification Network Module, which performs a binary classification to determine if the stream is 'Cover' (benign) or 'Stego' (contains hidden data). The remainder of this section provides a detailed explanation of each of these components, covering codewords correlation analysis, graph construction, and the network architecture.
\color{black}

\begin{algorithm}[t!]
\caption{Proposed steganalysis approach.}\label{algo1}
\begin{algorithmic}[1]
\INPUT VoIP speech stream
\OUTPUT Classification (Cover or Stego)

\vspace{10pt}
\State \textbf{Codewords Extraction:}
    \State Apply sliding detection window of size \(T\) frames to collect continuous packets
    \For{each frame $f_i$}
            \State Extract  3 LSF codewords $c_{1,i}, c_{2,i}, c_{3,i}$
    \EndFor
    \State Form QIS matrix:
     $C = \begin{bmatrix}
    c_{1,1} & c_{1,2} & \ldots & c_{1,T} \\
    c_{2,1} & c_{2,2} & \ldots & c_{2,T} \\
    c_{3,1} & c_{3,2} & \ldots & c_{3,T}
    \end{bmatrix}$

\vspace{10pt}
\State \textbf{Graph Construction: $G=(V,E)$}
    \State Create nodes: Each frame $f_i$ becomes a node $v_i$
    \State Set node features: Use codewords $c_{1,i}, c_{2,i}, c_{3,i}$ as 3D feature vector $x_i$ for each node $v_i$
    \State Create edges: Connect adjacent frames (nodes $v_i$ and $v_{i+1}$ ) with directed edges $v_j$
    \State Construct adjacency matrix $A$:\\
        \hspace{20pt} -- $A_{i,i}=0$ (no self-loops) \\
        \hspace{20pt} -- $A_{i,i+1}=1$ (connect successive nodes)
\vspace{10pt}
\State \textbf{GraphSAGE Network:}
    \State \textit{Graph Updating}:  
    \For{layer $k$: $1$ to  $K=3$}
        \For{node $v \in V$}
            \State Set initial attribute of $v:h_{v}^0=x_i$
            \State Sample neighborhood $N(v)$
            \State Aggregate information: \\ 
            \hspace{28pt} $h_{N(v)}^k = AGG_k(\{h_u^{k-1}, \forall u \in N(v)\})$
            \State Update node embedding: \\
            \hspace{28pt} $h_v^k = \mathcal{F}(W_k \cdot CONCAT(h_v^{k-1}, h_{N(v)}^k))$
        \EndFor
        \State Compute graph-level representation at layer $k$:
        \hspace{10pt} $z_G^k = \text{MeanPool}\left(\left\{h_v^k,v \in V\right\}\right) = \frac{1}{|V|}\sum_{v \in V}\left(\mathbf{h}_v^K\right)$
    \EndFor
    \State \textit{Hierarchical Pooling:}
        \State Compute final graph-level representation $z_G$ by summing all layer representations: 
        $z_G = \sum_k z_G^k$ 
\vspace{10pt}
\State \textbf{Classification:}
    \State Apply dropout to $z_G$
    \State Feed to fully connected layer
    \State Compute class probabilities using softmax
    \State Determine final classification (Cover or Stego)
\vspace{10pt}
\State \textbf{Training:}
    \State Use CrossEntropyLoss as the loss function
    \State Backpropagate and update model parameters.
\end{algorithmic}
\end{algorithm}

\subsection{Codewords correlation analysis}
As previously explained, the QIM steganography process involves making changes to the quantization of LSFs, affecting the LSF codewords denoted as $c_i$. Thus, all the necessary information for steganalysis is contained within these codewords. 

To perform steganalysis, a sliding detection window collects one or several continuous packets of compressed speech. This process allows for the construction of a sequence of quantized LSF codewords for analysis.
The G.723  and G.729 compression processes operate in frames with durations of 10ms and 30ms, respectively.
Assuming that the detection window size is denoted as T frames, and its frames can be represented as a set $F = [f_1, f_2, \ldots, f_T]$. The quantized LSF codewords, also known as the quantization index sequence (QIS), are expressed as $C = [c_{i,1}, c_{i,2}, \ldots, c_{i,T}]$, where $c_{i,j}$ is the $i^{th}$ codeword at frame $j$ in the sequence. For G.729 and G.723, where $i \in [1, 3]$, the QIS can be expressed in matrix form as:

\begin{equation}\label{eq:qis}
C = \begin{bmatrix}
c_{1,1} & c_{1,2} & \ldots & c_{1,T} \\
c_{2,1} & c_{2,2} & \ldots & c_{2,T} \\
c_{3,1} & c_{3,2} & \ldots & c_{3,T}
\end{bmatrix}
\end{equation}
In the case where all codewords are uncorrelated, their occurrences become independent. As a result, it is expressed as:
\begin{equation}
P(c_{i,j}, c_{k,l}) = P(c_{i,j}) \cdot P(c_{k,l}),
\end{equation}
where $P(c_{i,j}, c_{k,l})$  represents the joint probability of observing codewords $c_{i,j}$ and $c_{k,l}$,  and $P(c_{i,j})$,  $P(c_{k,l})$ denote their individual probabilities.
This expression is valid for all \(i\) and \(k\) in the range of 1 to 3, and \(j\) and \(l\) in the range of 1 to \(T\). 
Inequality between the two sides implies a correlation between \(c_{i,j}\) and \(c_{k,l}\).

Due to the \textcolor{black}{repeating patterns in human speech sounds (especially vowels and voiced consonants)}, the vocal signal exhibits stability over limited durations,  approximately the frame length. Consequently, the codewords exhibit correlations within the same frame.  Furthermore, speech signals have local periodicity, meaning that codewords have similar values in different frames. As a result, there are four types of inter-codeword correlations, namely: intra-frame correlation, successive frame correlation, cross-frame correlation, and cross-word correlation \cite{lin2018rnn}.  Local features are characterized by the first two correlations, while global ones are reflected by the latter two.

The QIM steganography process has a direct impact on the correlations between these codewords, thereby altering their statistical distribution. Earlier VoIP steganalysis approaches aimed to extract correlation features to uncover steganograms, as in   \cite{  li2017steganalysis,yang2018steganalysis}. 

In the proposed steganalysis approach, the power of GNNs, specifically the GraphSAGE model, is leveraged to capture these crucial correlations for steganalysis. GNNs have proven to be highly effective in modelling complex relationships and dependencies within graph-structured data, making them an ideal choice for understanding the intricate correlations between codewords in LBR VoIP streams. By representing the relationships between codewords as a graph and employing GraphSAGE, local and global connections can be effectively analyzed, thereby identifying patterns and changes induced by QIM steganography. This approach allows for the extraction of meaningful features that are instrumental in detecting steganographic content, thereby enhancing the accuracy and efficiency of the steganalysis process.

\subsection{Graph construction}
As GNNs require data in the form of a graph for processing during training, it is essential to transform the QIS matrix representation of the compressed speech stream into a graph structure. This transformation enables the GNN to leverage the inherent relationships and patterns within the speech data for effective steganalysis.

Taking the QIS matrix as input, the objective of this step is to construct a corresponding graph \(G = (V, E)\), where \(V = \{v_i\}_{i=1}^T\) is the set of \(T\) nodes, and \(E = \{e_j\}_{j=1}^{T-1}\) is the set of all edges between the nodes.
 
QIM-based steganography induces significant changes, principally in the correlation of adjacent frame codewords, i.e., edges between codewords of adjacent frames \cite{yang2022real, lin2018rnn, li2012detection, li2017steganalysis, yang2018steganalysis}. Consequently, each of the \(T\) frames in the speech signal is represented as a node, forming a graph \(G\) composed of \(T\) nodes:

\begin{itemize}
    \item Each node \(v_i\) is assigned a feature vector \(x_i \in \mathbb{R}^3\) that encapsulates the three codewords from the corresponding frame: \(x_i = [c_{1,i}, c_{2,i}, c_{3,i}]\).

    \item The edges \(e_j\), which link the nodes \(v_i\) and \(v_{i+1}\), are directional and symbolize the transitions between two successive frames. \textcolor{black}{This choice of directed edges is crucial for capturing the temporal sequence and dependencies inherent in speech signals, allowing the model to learn how information flows and changes over time, particularly between consecutive frames. This design choice aligns with prior studies in speech processing, where directed relationships are commonly employed to model temporal dependencies and improve learning performance \cite{shirian2021compact, li2023speech}.}
\end{itemize}

The graph generated from these procedures is a directed acyclic graph (DAG) used to represent the temporal dependencies between frames in the compressed speech.  Due to its directed configuration, the graph ensures that information flows in one direction, mirroring the progression of time in speech signals. This unidirectional flow is essential for capturing the temporal dependencies that may be altered by steganographic embedding. The acyclic property of the graph prevents loops, ensuring that the model doesn't incorrectly assume circular dependencies between frames. This is crucial for maintaining the integrity of the temporal sequence and avoiding spurious correlations. 

\begin{figure}[t]
    \centering
    \includegraphics[scale=0.57]{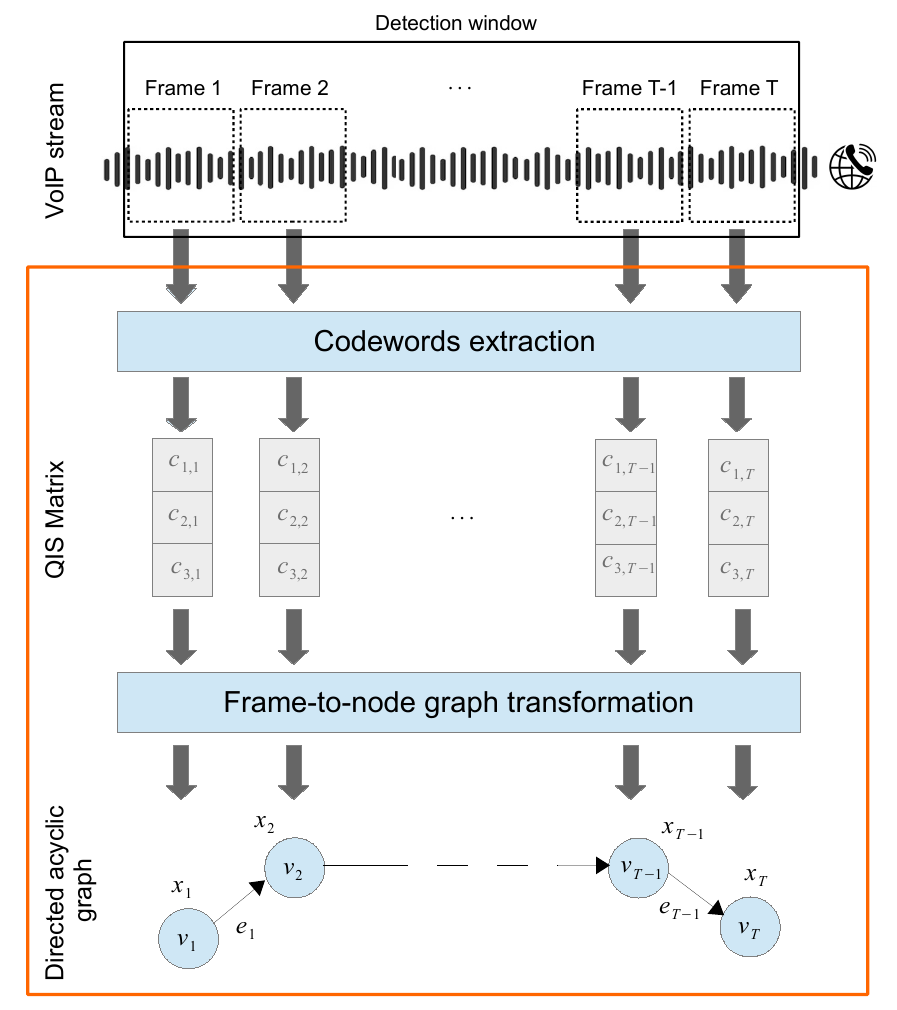}
    \caption{Speech-to-graph construction process.}
    \label{fig:graph_const}
\end{figure}

The proposed graph construction process is depicted in Figure \ref{fig:graph_const}.

The adjacency matrix of \(G\) is indicated by  \(A \in \mathbb{R}^{T \times T}\), with \((A)_{ij}\) indicating the connection weight between nodes \(v_i\) and \(v_j\), as illustrated in Equation \ref{eq:adjencymatrix}. The off-diagonal elements,  signify the presence of edges between nodes, with a value of 1 indicating their connectivity. The diagonal elements, by contrast, are all set to 0, as nodes are not self-connected in this context.

\begin{equation}\label{eq:adjencymatrix}
A = \begin{bmatrix}
0 & 1 & 0 & \cdots & 0 \\
1 & 0 & 1 & \cdots & 0 \\
0 & 1 & 0 & \cdots & 0 \\
\vdots & \vdots & \vdots & \ddots & \vdots \\
0 & 0 & \cdots & 1 & 0
\end{bmatrix}
\end{equation}

The construction of a simple graph structure in this context is well-justified due to several essential factors. First and foremost, simplicity in the graph structure reduces the computational complexity and resource requirements during the training phase, particularly in scenarios involving large datasets. This efficiency is valuable not only during training but also in the testing phases, as it allows for the rapid analysis of a significant number of speech segments without excessive resource consumption. Moreover, the straightforward graph structure aligns perfectly with the nature of the information under examination. In QIM-affected speech data steganalysis, the critical alterations predominantly occur in the transitions between adjacent frames, as already mentioned. Consequently, this approach effectively captures these local temporal dependencies. Furthermore, despite the graph's simplicity, the GraphSAGE model excels at feature extraction, effectively aggregating local neighbourhood information to reveal the vital temporal dependencies critical for accurate steganalysis.

\subsection{Network architecture}
Given a collection of graphs \(\{G_1, \ldots, G_N\}\), derived from speech samples, along with their corresponding true labels \(\{y_1, \ldots, y_N\}\), the objective is centred around graph classification. The aim is to differentiate between graphs corresponding to cover data and those indicating the presence of QIM steganography.
 
\begin{figure*}
    \centering
    \includegraphics[scale=0.57]{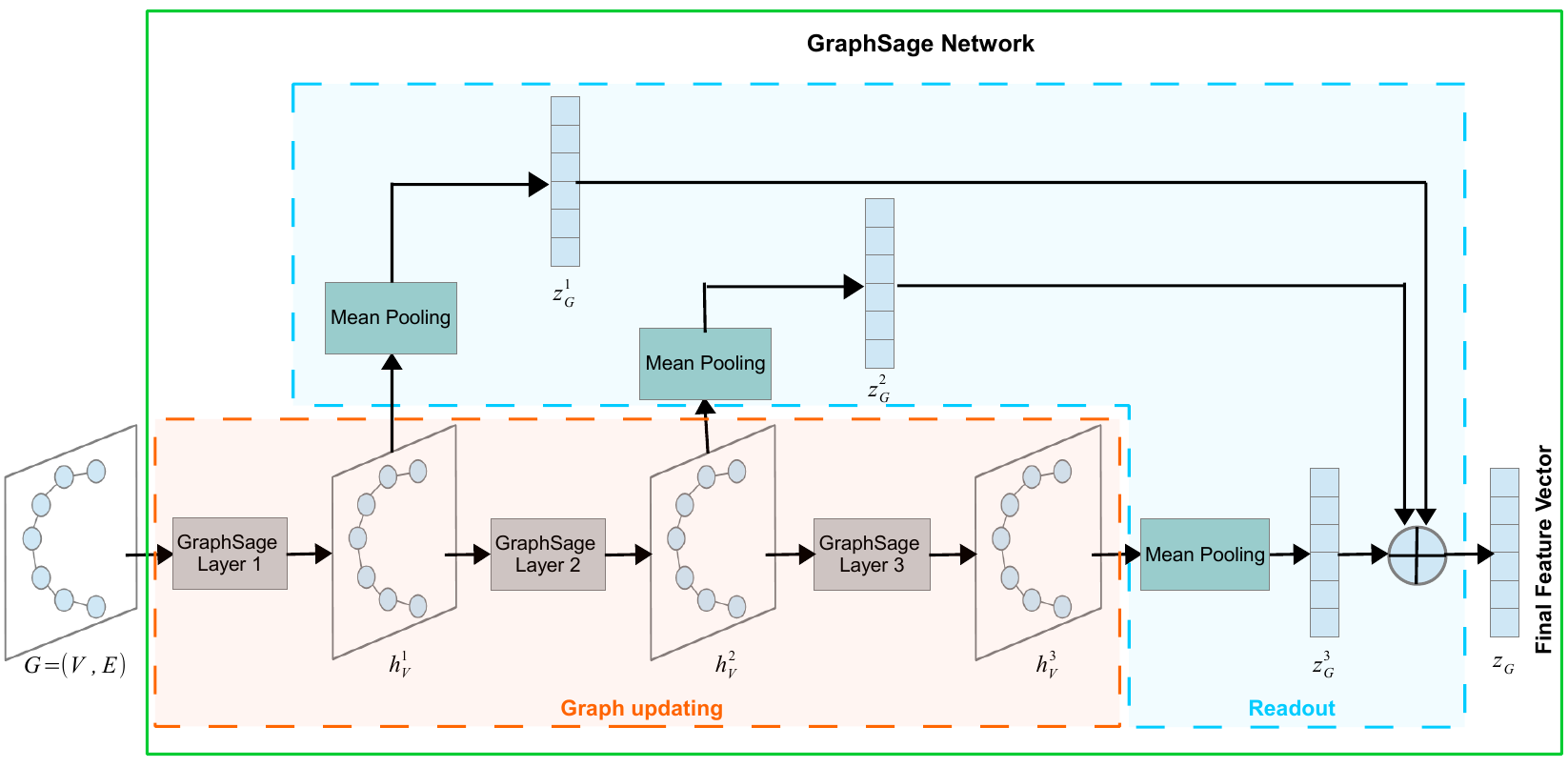}
    \caption{The proposed GraphSAGE-based network architecture.}
    \label{fig:graphSageNet}
\end{figure*}

To accomplish this, the generated graphs are passed through a GraphSAGE network for representation learning.  This process produces a feature vector that captures the most relevant information indicating the presence of QIM steganography, which is then used in the final classification step.
The network architecture, as depicted in Figure  \ref{fig:graphSageNet}, comprises two sub-networks: \textit{GraphSAGE Network} and \textit{Classification Network}.

\subsubsection{GraphSAGE network}
In this architectural framework, the GraphSAGE Network consists of two key phases: \textit{graph updating} and \textit{readout (pooling)}. The objective of this network is to extract a vector of the most relevant features for steganalysis.

During the graph updating phase, the node embeddings within the graph undergo a series of operations aimed at refining and modifying them. This process is crucial for extracting essential features in steganalysis. In our architecture, three GraphSAGE convolution layers are employed to progressively extract more abstract features from the input graph data.
The GraphSAGE algorithm, suggested by Hamilton et al. \cite{hamilton2017inductive}, is instrumental in this process.

Starting with the input graph \(G = (V, E)\) including all feature vectors corresponding to the graph nodes \(\{x_v, \forall v \in V\}\), the GraphSAGE algorithm is determined by how many graph convolutional layers \(K\) are employed, which in our design is set to 3. This determines the number of hops used for aggregating node information. Furthermore, differentiable aggregator functions \(AGG_k, \forall k \in \{1, \ldots, K\}\) are employed to combine information from neighbouring nodes. At each iteration, the algorithm collects information from a node's neighbours, their neighbours, and so on. 

Each iteration involves sampling nearby nodes and summarizing their features into a consolidated vector. At the \(k^{th}\) layer, the combined features for a node \(v\) based on the sampled neighborhood \(N(v)\),  \(h_{N(v)}^k\),  is described in Equation \ref{eq:graphsage1}:

\begin{equation}\label{eq:graphsage1}
h_{N(v)}^k = AGG_k(\{h_u^{k-1}, \forall u \in N(v)\}),
\end{equation}

\noindent where \(h_u^{k-1}\) corresponds to the output of node 
\(u\) from the previous layer.  The embeddings of all nodes \(u\) within the neighbourhood of node \(v\) are collectively merged to form the embedding of node \(v\) at layer \(k\). In Equation \ref{eq:graphsage1}, various aggregation functions such as mean, pooling, graph convolution, or LSTM can be applied. Within the scope of our model, the LSTM architecture achieved superior performance, as proved later in experiments.
\color{black}
The LSTM aggregator processes the set of neighbor representations as a sequence through an LSTM. For each node $v$:
\begin{equation}
    h_{N(v)}^k = \text{LSTM}(h_{u_1}^{k-1}, h_{u_2}^{k-1}, \dots, h_{u_m}^{k-1})
    \label{eq:lstm_agg_sequence}
\end{equation}
where $(u_1, u_2, \dots, u_m)$ is a random permutation of $N(v)$. The final hidden state of the LSTM sequence processing is used as the aggregated vector $h_{N(v)}^k$.
\color{black}

The neighbourhood information, represented by the aggregated embeddings \(h_{N(v)}^k\), is combined with the previous layer's embedding of the node \(v\), \(h_v^{k-1}\), through concatenation. This concatenated vector is then transformed by the trainable weight matrix \(W_k\) and passed through a non-linear activation function \(\mathcal{F}\) (e.g., rectified linear unit (ReLU)). The outcome yields the updated node representation at layer \(k\), as described in Equation~\ref{eq:graphsage2}:

\begin{equation}\label{eq:graphsage2}
h_v^k = \mathcal{F}\left(W_k \cdot \text{CONCAT}(h_v^{k-1}, h_{N(v)}^k)\right)
\end{equation}

\begin{figure*}
    \centering
    \includegraphics[scale=0.6]{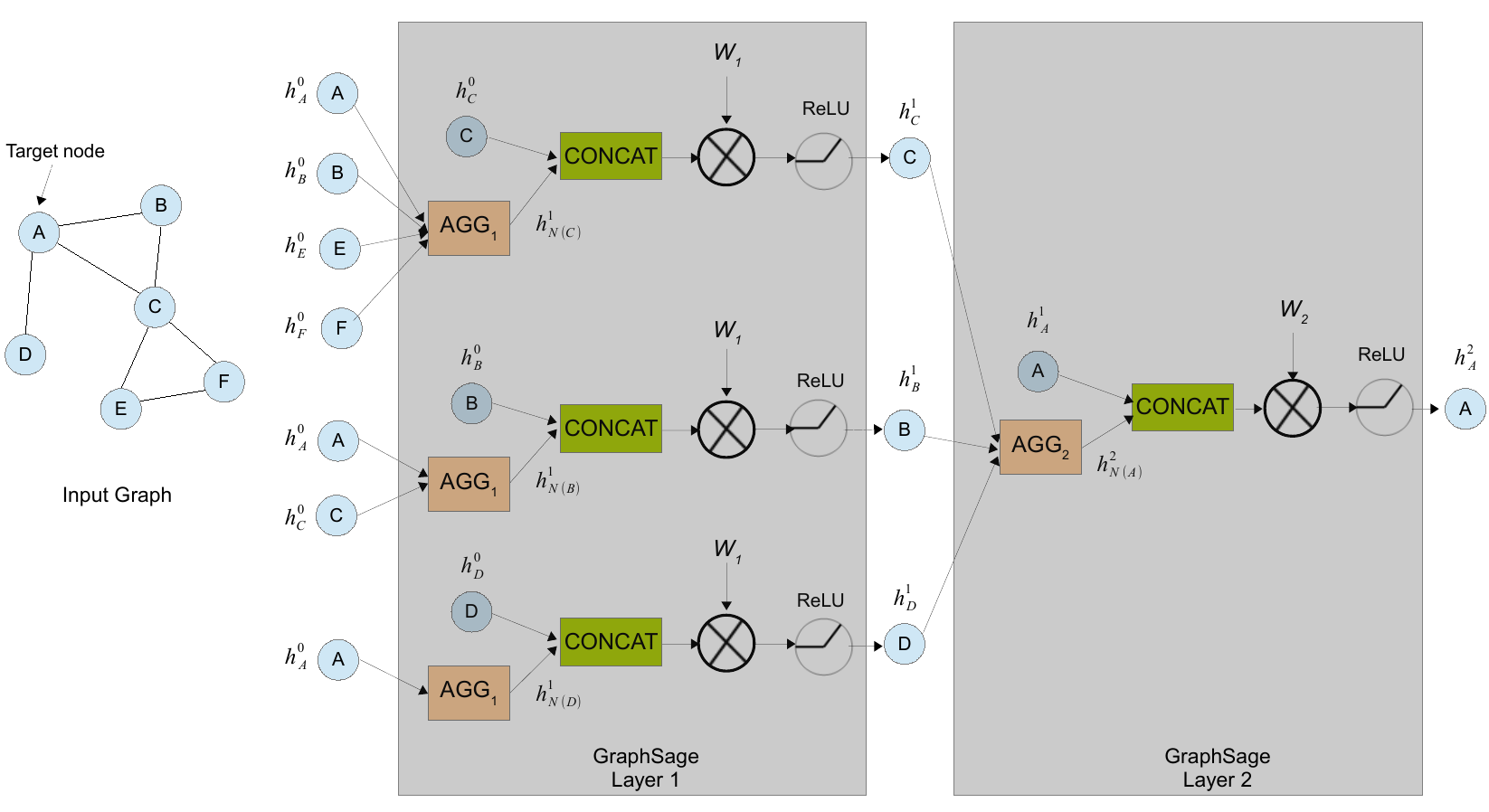}
    \caption{A two-layer node embedding process. \(h_v^0\) denotes the initial attribute of node \(v\). In this illustration, node embedding is shown for the target node \(A\). Notably, the embedding of other nodes occurs in parallel.}
    \label{fig:GraphSageLayers}
\end{figure*}

The final representation of node \(v\), denoted as \(z_v\), is obtained from the node's embedding at the last layer \(K\), i.e., \(h_v^K\), as formulated in Equation~\ref{eq:graphsage3}. The procedure of node embedding through two GraphSAGE layers is depicted in Figure \ref{fig:GraphSageLayers}.

\begin{equation}\label{eq:graphsage3}
z_v = h_v^K, \forall v \in V
\end{equation}

The objective is graph classification, where graphs are aimed to be categorized into two classes: stego and cover. This task differs from the more common objective of classifying nodes or edges within a graph. To achieve this, a graph-level representation \(z_G \in \mathbb{R}^Q\)  needs to be generated from the node embeddings. This is generally accomplished by applying a pooling operation over the final-layer node embeddings \(z_v\). Instead of this, a hierarchical pooling approach has been introduced 
by pooling the node-level embeddings at each layer \(h_v^k, \forall k \in \{1, \ldots, K\}\). 
This approach allows for capturing features at different levels of abstraction. By aggregating information hierarchically, a comprehensive graph-level representation is created 
that encompasses both fine-grained details and high-level patterns. This proves particularly valuable in steganalysis tasks, where subtle changes in different parts of the graph can collectively indicate the presence of hidden information. 
Common options for pooling methods in graph analysis include mean pooling, max pooling, and sum pooling. In our QIM steganalysis context, the use of mean pooling was found to be especially effective, as it averages the features at each layer,  enabling the capture of collective characteristics crucial for QIM steganography detection.
The graph-level representation at each layer \(k\), \(z_G^k\),  is defined as in Equation \ref{eq:pooling1}:
\begin{equation}\label{eq:pooling1}
z_G^k   = \text{MeanPool}\left(\left\{h_v^k, v \in V\right\}\right) = \frac{1}{|V|}\sum_{v \in V}\left(\mathbf{h}_v^K\right),
\quad \forall k \in \{1, \ldots, K\}
\end{equation}
where MeanPool is the pooling operation that averages the node representations. \textcolor{black}{Mean pooling preserves the collective statistics and the overall distribution of features across nodes, which is crucial for detecting subtle patterns introduced by QIM steganography in compressed speech streams. In contrast, max pooling may discard valuable information by focusing only on the strongest activation, making it less effective when the signal differences are minimal and widely distributed. This observation is supported by the results presented in Section 5.4, which show that using max pooling instead of mean pooling leads to a notable drop in detection accuracy.}

The final vector representing the entire graph, \(z_G\), is constructed by summing all the hierarchically pooled representations \(z_G^k\), as expressed in Equation \ref{eq:pooling2}.

\begin{equation}\label{eq:pooling2}
z_G = \sum_k^K z_G^k
\end{equation}

This global representation encapsulates the most relevant features for steganalysis. Additionally, it acts as the input for the subsequent classification layer, allowing the model to make predictions about the presence of steganography in the input data.

\subsubsection{Classification task}

After obtaining the comprehensive graph-level representation \( z_G \), the model proceeds to the classification stage. The primary purpose of this stage is to differentiate between regular speech (cover data) and speech samples with embedded QIM steganography (stego speech). The classification incorporates two key layers: a dropout layer and a linear classification layer, referred to as the fully connected layer.

The first step is to apply dropout to \( z_G \) to prevent overfitting. To prevent overfitting, dropout randomly sets a portion \(p\) of input activations to zero during training updates. The dropout layer can be formulated as:

\begin{equation}
z_{D} = \mathcal{D}(z_G, p),
\end{equation}
where \(\mathcal{D}\) denotes the dropout function, and \(z_{D}\) is the resulting vector after applying this fucntion. 

The result from the dropout layer is subsequently passed to the fully connected layer to perform the final classification. The latter layer maps the representation to the output space for binary classification, with output neurons equal to the number of target classes (two in our case: "cover" and "stego"). Mathematically, the final classification $\hat{y}$ can be expressed as follows:

\begin{equation}
    \hat{y} = W_{FC} z_{D} + b_{FC},
\end{equation}

\noindent where \( W_{FC} \) and \( b_{FC} \) represent the weights and biases of the fully connected layer, respectively. The output $\hat{y}$ corresponds to the raw logits for each class.

The model employs CrossEntropyLoss as its loss function for the binary classification task, which combines a softmax activation and a negative log-likelihood loss. The function is defined as:

\begin{equation}
\small
\text{Loss} =  -\frac{1}{N} \sum_{i=1}^{N} \left[ y_i \log\left(\frac{\exp(\hat{y_i})}{\sum_{j} \exp(\hat{y_j})}\right) \right.
 \left. + (1 - y_i) \log\left(1 - \frac{\exp(\hat{y_i})}{\sum_{j} \exp(\hat{y_j})}\right) \right]
\end{equation}

\noindent where Loss is the loss value to be minimized, \( y_i \) is the true label, \( \hat{y_i} \) are the logits for the \(i\)-th sample, and  \(N\) is the number of samples in the dataset. \(\sum_{j} \exp(\hat{y_j})\) represents the sum of the exponentials of logits across all classes, used for normalization in the softmax function.

\textcolor{black}{The proposed architecture, consisting of graph construction, the GraphSAGE network, and the classification module, is implemented and evaluated under various experimental configurations, as detailed in the next section.}

\section{Experiments and discussion}\label{section5}
The efficacy of the suggested steganalysis technique is assessed through various experiments, focusing on factors such as the embedding rate, duration of the speech sample, and time consumption.

\subsection{Experiment setup}

The experiments were conducted using a dataset described in \cite{lin2018rnn} and available on the  Github\footnote{\url{https://github.com/fjxmlzn/RNN-SM}} platform. This dataset includes 72 hours of English speech and 41 hours of Chinese speech, totalling 320 recordings in 16-bit PCM format. The speech samples were sourced from the internet and included contributions from various male and female speakers.
In our methodology, the G.729A LBR speech codec was employed
to encode the original speech samples, forming the basis of our cover speech dataset. To create the stego samples, the CNV-QIM \cite{xiao2008approach} steganography method was used, inserting secret bits during the G.729A encoding phase.

To assess the robustness of our detection technique under different conditions, stego samples were generated at varying embedding rates ranging from 20\% to 100\% with a step size of 20\%. Moreover, the algorithm's ability to detect samples of varied lengths was assessed by segmenting entries in both the cover and stego speech datasets into sample lengths \(L_s\) of 0.5s, 1s, 3s, 5s, 7s, and 10s. For each test, samples were chosen from the stego and cover datasets based on the designated embedding rate and duration. For example, for the training set with 10s segments and an embedding rate of 100\%, there were 10,552 cover samples and 10,552 stego samples. \textcolor{black}{The speech data were partitioned into three subsets — 70\% for training, 15\% for validation, and 15\% for testing — using a randomized split while maintaining a 1:1 cover-to-stego ratio within each subset to ensure balanced and unbiased evaluation.}
The validation subset was employed to fine-tune the model's parameters, whereas the testing portion served to assess the model’s effectiveness.
\textcolor{black}{A comprehensive summary of the dataset characteristics is provided in Table \ref{tab:dataset_summary}.}

\begin{figure*}[t]
    \centering
    \includegraphics[scale=0.55]{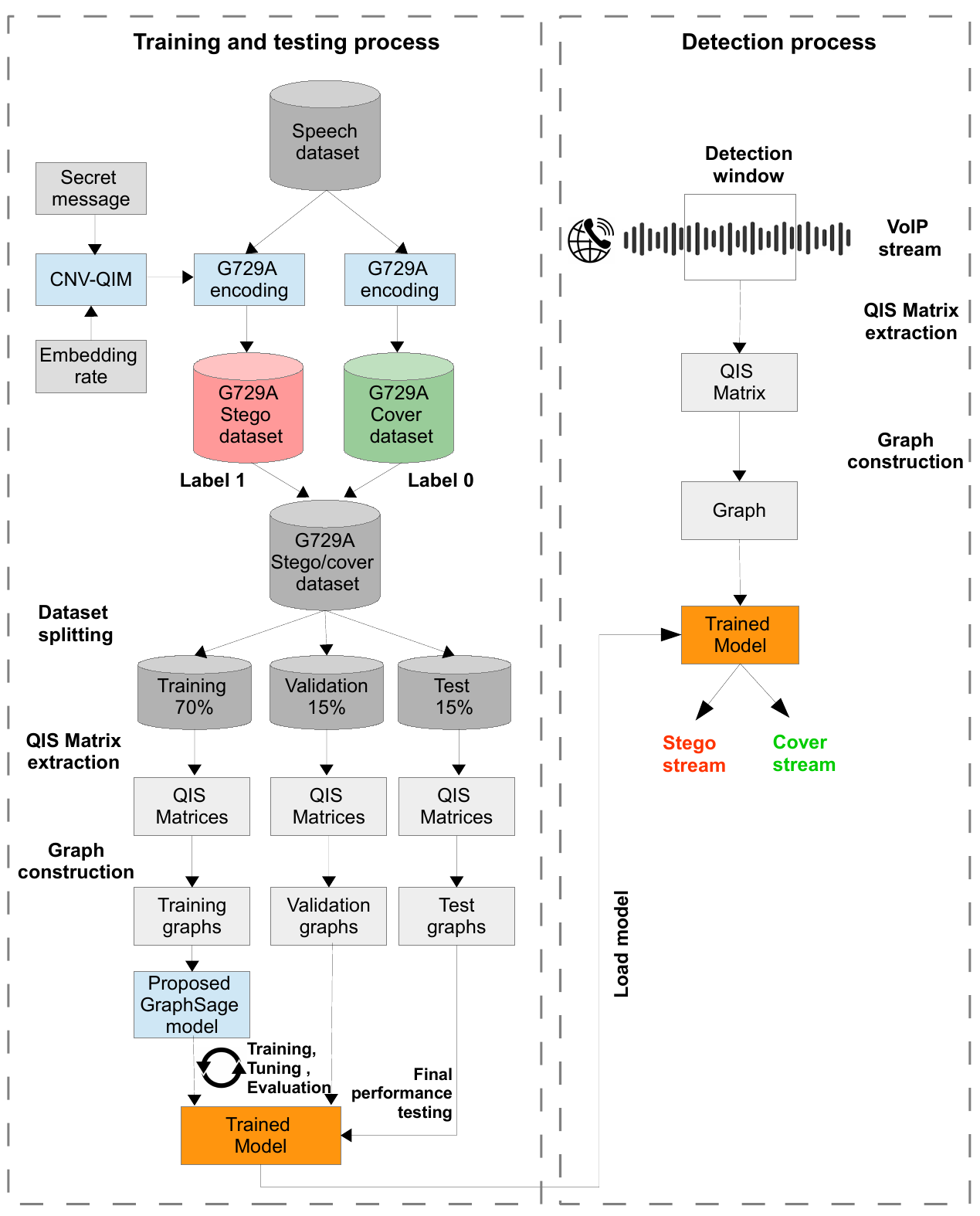}
    \caption{Process of training, testing and detection.}
    \label{fig:TrainTestProcess}
\end{figure*}

The process of training and testing our model is depicted in Figure \ref{fig:TrainTestProcess} (left-hand process).

\begin{table}[t!]
\color{black}
\centering
\caption{\textcolor{black}{Dataset description}}
\begin{tabular}{lp{4.3cm}}
\toprule
\textbf{Feature} & \textbf{Description} \\
\midrule
Total Duration & 113 hours (72 hours English, 41 hours Chinese) \\
Number of Recordings & 320 recordings in 16-bit PCM format \\
Speakers & Various male and female speakers (sourced from the Internet) \\
Codec & G.729A LBR speech codec \\
Steganography Method & CNV-QIM embedding during G.729A encoding \\
Embedding Rates Tested & 20\%, 40\%, 60\%, 80\%, 100\% \\
Segment Lengths ($L_s$) & 0.5s, 1s, 3s, 5s, 7s, 10s \\
Train/Validation/Test Split & 70\% / 15\% / 15\% (randomized) \\
Cover/Stego Ratio & 1:1 in each subset \\
\bottomrule
\end{tabular}
\label{tab:dataset_summary}
\color{black}
\end{table}

The proposed steganalysis model was trained on the Kaggle platform using a P100 GPU. The model was implemented with the PyTorch and PyTorch Geometric frameworks, using Adam as the optimizer, configured with a learning rate of 0.003  for efficient convergence. A batch size of 32 was employed during data preprocessing and model training, spanning 150 epochs to improve the proposed GNN-based steganalysis model's performance. \textcolor{black}{The key hyperparameters used during training and model configuration are summarized in Table \ref{tab:hyperparameters}. These parameters were selected empirically through iterative experimentation to achieve the best detection accuracy and generalization performance} 

\begin{table}[]
\color{black}
\centering
\caption{\textcolor{black}{Summary of hyperparameters used for model training}}
\label{tab:hyperparameters}
\begin{tabular}{ll}
\toprule
\textbf{Hyperparameter} & \textbf{Value} \\ \midrule
Optimizer & Adam \\ 
Learning rate & 0.003 \\ 
Batch size & 32 \\ 
Number of epochs & 150 \\ 
Dropout rate & 0.3 \\ 
GNN layers & GraphSAGE \\ 
Number of layers& 3 \\ 
Hidden chanels & 64 \\ 
Aggregation function &  LSTM \\ 
Pooling Method &  Mean \\ 
Activation function & ReLU \\ 
Loss function & Binary Cross-Entropy \\ \bottomrule
\end{tabular}
\color{black}
\end{table}

\subsection{Evaluation metrics}

The proposed model is assessed using key performance indicators, namely \textit{detection accuracy}, \textcolor{black}{\textit{precision}, \textit{recall}, \textit{F1-score}, \textit{detection time}, and \textit{computational complexity}.}

\color{black}
The classification-related measures (accuracy, precision, recall, and F1-score) are computed based on the results of binary classification, which are summarized through four categories in the confusion matrix:

\begin{itemize}
    \item \textbf{True Positive (TP):} Instances that are actually stego and are correctly predicted as stego.
    \item \textbf{True Negative (TN):} Instances that are actually cover and are correctly predicted as cover.
    \item \textbf{False Positive (FP):} Instances that are actually cover but are incorrectly predicted as stego.
    \item \textbf{False Negative (FN):} Instances that are actually stego but are incorrectly predicted as cover.
\end{itemize}

The evaluation metrics are defined as follows:

\noindent\textbf{Detection accuracy:} This metric measures the proportion of correctly identified instances of steganography compared to the overall instances. It is calculated as:

\begin{equation}
\text{Accuracy} = \frac{\text{TP + TN}}{\text{TP + TN + FP + FN}}
\end{equation}
Detection accuracy is a key metric in our case as the dataset is evenly distributed, containing the same number of positive (stego) and negative (cover) samples. In a balanced dataset, the number of instances in each class is approximately equal, making accuracy a reliable indicator of the model's performance. High detection accuracy reflects the model’s ability to effectively distinguish between stego and non-stego signals, making it a suitable metric for evaluating performance in this balanced setting. 

\noindent\textbf{Precision:} Measures the accuracy of positive predictions made by the model. It quantifies the proportion of correctly identified stego samples out of all samples that the model predicted as stego. High precision indicates a low rate of false positives, which is important to avoid incorrectly flagging legitimate communication as covert. It is calculated as:

\begin{equation}
\text{Precision} = \frac{\text{TP}}{\text{TP + FP}}
\end{equation}

\noindent\textbf{Recall (Sensitivity):} Measures the model's ability to find all positive instances. It quantifies the proportion of correctly identified stego samples out of all actual stego samples present in the dataset. High recall indicates a low rate of false negatives, which is crucial for not missing any covert communications. It is calculated as:
\begin{equation}
\text{Recall} = \frac{\text{TP}}{\text{TP + FN}}
\end{equation}

\noindent\textbf{F1-score:} The F1-score is the harmonic mean of Precision and Recall, providing a single metric that balances both. It is particularly useful when dealing with imbalanced datasets (though less critical for our balanced dataset) or when both false positives and false negatives carry significant costs. It is calculated as:
\begin{equation}
\text{F1-score} = 2 \times \frac{\text{Precision} \times \text{Recall}}{\text{Precision + Recall}} = \frac{2 \times \text{TP}}{2 \times \text{TP + FP + FN}}
\end{equation}
\color{black}

\noindent \textbf{Detection time (DT):} Evaluates the time required for the model to process and classify an input sample. 
DT is critical for assessing the efficiency of the model, especially in real-time or online steganalysis scenarios. A shorter DT reflects the model's capability to perform analysis swiftly, making it suitable for time-sensitive applications. \textcolor{black}{Since the training phase is performed offline and does not impact real-time detection performance, which is the primary objectif of our work, we focus solely on the detection time for evaluating runtime efficiency.}

\color{black}
\noindent \textbf{Computational Complexity:} This metric provides insight into the model's inherent resource requirements, indirectly indicating its memory footprint, model size, and raw computational load. It is quantified in terms of:
\begin{itemize}
\item \textit{Parameters:} The total number of adjustable values within the model's architecture or its underlying algorithms, which are determined during the training phase. For neural networks, these are the trainable weights and biases; for traditional machine learning methods, these may include coefficients of statistical models or support vectors of classifiers. This directly correlates with model size and memory usage.
\item \textit{Floating Point Operations (FLOPs):} The number of arithmetic operations (additions, multiplications, etc.) required for a single inference pass. FLOPs indicate the raw computational intensity of the model, which influences execution time.
\end{itemize}
\color{black}

\subsection{Detection performance}

This section presents the evaluation results of the proposed approach under various scenarios, encompassing different sample lengths and embedding rates. The assessment begins by examining the impact of embedding rates on the system. In real-world scenarios, attackers often disperse secret information over an extended period to lower the average embedding rate, intending to enhance communication concealment. Consequently, effectively detecting steganographic VoIP signals at low embedding rates remains an open and difficult problem in the research community.

\setlength{\leftmargini}{15pt}
\begin{enumerate}[label=(\alph*),]
\item \textbf{Performance results under varying embedding rates:} 

A preliminary assessment of the degree of difficulty faced by the proposed GNN scheme in uncovering the CNV-QIM steganographic method, particularly at low embedding rates, was conducted using statistical analysis by comparing cover and stego datasets.  Figure \ref{fig:boxplot} presents this analysis using boxplots, contrasting cover and stego datasets across various embedding rates. The results demonstrate that CNV-QIM maintains consistent mean values across different scenarios, with only slight variations in the first (Q1) and third (Q3) quartiles as embedding rates increase. This stability indicates that CNV-QIM is a challenging steganographic method.
\color{black} The evaluation of the proposed steganalysis technique under different embedding rates for 10-second sample segments, as presented in Table \ref{tab:AccuER}, reveals nuanced performance characteristics. Notably, \textcolor{black}{the detection capability, as measured across all metrics (Accuracy, Precision, Recall, F1-score),} exhibits a slight decreasing trend with lower embedding rates. This is attributed to the fact that higher embedding rates induce more pronounced changes in codewords, providing a more discernible pattern for steganalysis. It is noteworthy that even at a low embedding rate of 20\%, our approach maintains high accuracy, registering at 95.17\%\textcolor{black}{, and demonstrates strong, balanced performance across the other metrics, with Precision at 95.43\%, Recall at 94.95\%, F1-score at 95.19\%, and AUC at 95.20\%.} {This high level of accuracy in a low-payload scenario is particularly significant, as attackers frequently use low embedding rates to minimize statistical disturbances and evade detection. The model's ability to maintain such robust performance demonstrates its high sensitivity to even the most subtle steganographic artifacts.}

To further illustrate this robust performance in the challenging low-embedding-rate scenario, Figure \ref{fig:confusion_matrix} presents the detailed confusion matrix for the 20\% embedding rate. Out of a total of 3166 test samples (comprising 1583 actual stego and 1583 actual cover samples), the model successfully identified 1503 TP (correctly classified stego) and 1511 TN (correctly classified cover). Critically, the false classifications were minimal, with only 72 FP (cover incorrectly flagged as stego) and 80 FN (stego samples missed). {This granular breakdown is crucial for practical applications, as it demonstrates the model's ability to strike a critical balance: minimizing false alarms (FP), which could disrupt legitimate communication, while simultaneously minimizing missed detections (FN), which would allow covert channels to go unnoticed.} The proposed steganalysis approach consistently achieves high detection rates despite the inherent challenge posed by the resilience of CNV-QIM, showcasing its strong discriminative power even under subtle embedding. {This demonstrates that the GNN-based architecture successfully learns to distinguish the subtle, distributed changes introduced by steganography from the natural variations in cover speech.}

\begin{figure}
    \centering
    \includegraphics[scale=0.65]{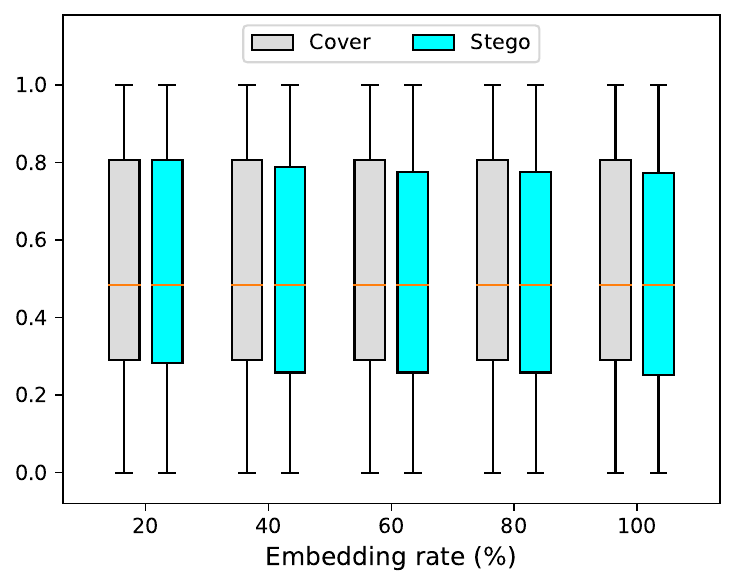}
    
    \caption{Statistical analysis of cover and stego speeches at different embedding rates.}
    \label{fig:boxplot}
\end{figure}

\begin{table}
    \centering
    \caption{\textcolor{black}{Detection performance metrics} across varied embedding rates for 10s samples.}
    \label{tab:AccuER}
    \begin{tabular}{p{2.9cm}lllll}
         \toprule
         Embedding rate (\%)& 20 & 40 & 60 &80 & 100 \\ \midrule
         Accuracy (\%)& 95.17 & 99.68 &99.94& 100 & 100 \\
         \textcolor{black}{Precision (\%)}& \textcolor{black}{95.43} & \textcolor{black}{99.68} & \textcolor{black}{99.94} & \textcolor{black}{100} & \textcolor{black}{100} \\
         \textcolor{black}{Recall (\%)}& \textcolor{black}{94.95} & \textcolor{black}{99.68} & \textcolor{black}{99.94} & \textcolor{black}{100} & \textcolor{black}{100} \\
         \textcolor{black}{F1-score (\%)}& \textcolor{black}{95.19} & \textcolor{black}{99.68} & \textcolor{black}{99.94} & \textcolor{black}{100} & \textcolor{black}{100} \\
         \bottomrule
    \end{tabular}
\end{table}

\begin{figure}
    \centering
    \includegraphics[scale=0.6]{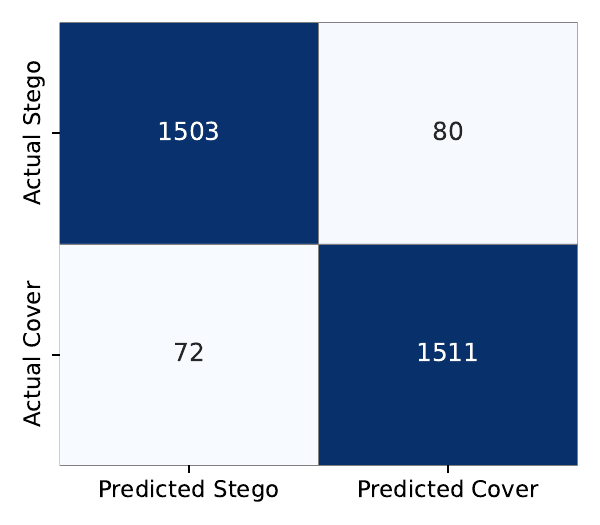}
    
    \caption{\textcolor{black}{Confusion matrix under 20\% embedding rate.}}
    \label{fig:confusion_matrix}
\end{figure}

\item \textbf{Performance results under varying speech segment lengths:} In VoIP steganalysis, the length of speech segments is also a crucial factor influencing detection accuracy. A successful approach must achieve a sufficiently high level of accuracy within a limited time detection window. Table \ref{tab:AccuSpeechLengh} explores the impact of sample length \textcolor{black}{on various detection performance metrics}, focusing on a fixed embedding rate of 100\%. 
The findings reveal that longer sample lengths lead to a steady and significant improvement \textcolor{black}{across all performance metrics: Accuracy, Precision, Recall, and F1-score. This improvement is rapid, with performance reaching near-perfect levels (e.g., 99.97\% accuracy at 3 seconds) and achieving a perfect 100\% across all metrics from 7-second samples onwards. {This trend is directly linked to the nature of our GNN-based model; a longer speech segment translates to a larger graph with more nodes and edges. This allows the GraphSAGE architecture to aggregate information over a wider neighborhood, resulting in more stable and representative graph-level embeddings that make the distinction between cover and stego patterns far more pronounced.} Crucially, even with short 0.5-second segments, the approach we propose attains exceptionally high detection performance, with Accuracy, Precision, Recall, and F1-score all exceeding 98\%. {This is a critical finding for practical deployment, as real-time monitoring systems must often make decisions based on individual or fragmented data packets. Furthermore, attackers may intentionally disperse hidden data in short, disjointed bursts to evade detection systems that require longer samples for analysis.} The remarkable consistency across these metrics for each sample length highlights the model's robust and balanced performance, demonstrating its ability to minimize both false positives and false negatives. The method therefore enables effective detection of QIM steganography even for very short segments of monitored VoIP traffic, which is vital for real-world scenarios where covert data might be dispersed in brief bursts.}


\begin{table}[h!]
    \centering
    \caption{\textcolor{black}{Detection performance metrics} across varied sample lengths  at 100\% embedding rate.}
    \label{tab:AccuSpeechLengh}
    \begin{tabular}{m{1.95cm}m{0.7cm}m{0.7cm}m{0.7cm}m{0.7cm}m{0.7cm}m{0.7cm}}
    \toprule
    Sample length (s)&0.5 & 1 & 3  & 5 & 7 & 10\\\midrule
    Accuracy (\%)&98.26 &99.47  & 99.97 & 99.98 & 100 & 100\\
    \textcolor{black}{Precision (\%)} & \textcolor{black}{98.27} & \textcolor{black}{99.43} & \textcolor{black}{99.97} & \textcolor{black}{99.98} & \textcolor{black}{100} & \textcolor{black}{100} \\
    \textcolor{black}{Recall (\%)} & \textcolor{black}{98.27} & \textcolor{black}{99.49} & \textcolor{black}{99.97} & \textcolor{black}{99.98} & \textcolor{black}{100} & \textcolor{black}{100} \\
    \textcolor{black}{F1-score (\%)} & \textcolor{black}{98.24} & \textcolor{black}{99.46} & \textcolor{black}{99.97} & \textcolor{black}{99.98} & \textcolor{black}{100} & \textcolor{black}{100} \\
    \bottomrule
\end{tabular}
\end{table}


\item \textbf{Performance results with varying speech segment lengths and embedding rate:} Figure \ref{fig:AccuEmRateSpeechLentgh} provides the results of detection accuracy with varying embedding rates for sample segments of different lengths. The results reaffirm the method's reliability, with a general upward trend in accuracy as the embedding rate and sample length increase. {This trend is expected, as longer samples and higher embedding rates provide the model with more data and a stronger, less ambiguous signal, respectively, making the detection task easier.} It is evident that achieving accurate detection under low embedding rates with short segments poses a challenge {for any steganalysis system. This scenario, representing the most difficult detection conditions due to minimal available information, can be considered a benchmark for model robustness. Our approach maintains an acceptable accuracy exceeding 70\% even in these worst-case conditions, which, rather than being an "Achilles' heel," demonstrates that the GNN is still able to capture a discriminative signal from extremely sparse and noisy data, highlighting its effectiveness.}

\begin{figure}
    \centering
    \includegraphics[scale=0.47]{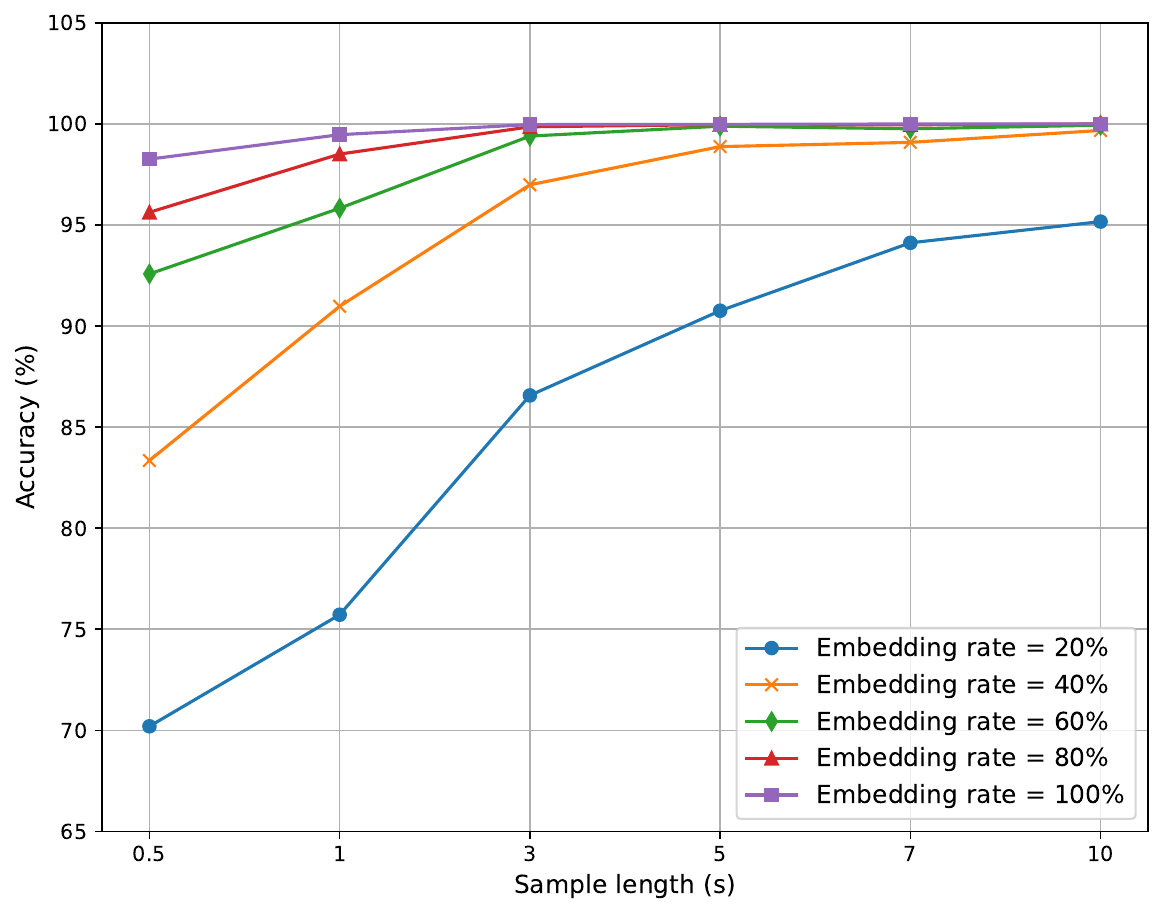}
    \caption{Detection accuracy across varying embedding rates and sample length.}
    \label{fig:AccuEmRateSpeechLentgh}
\end{figure}

\end{enumerate}

\begin{figure*}[t!]
    \centering
\includegraphics[scale=0.55]{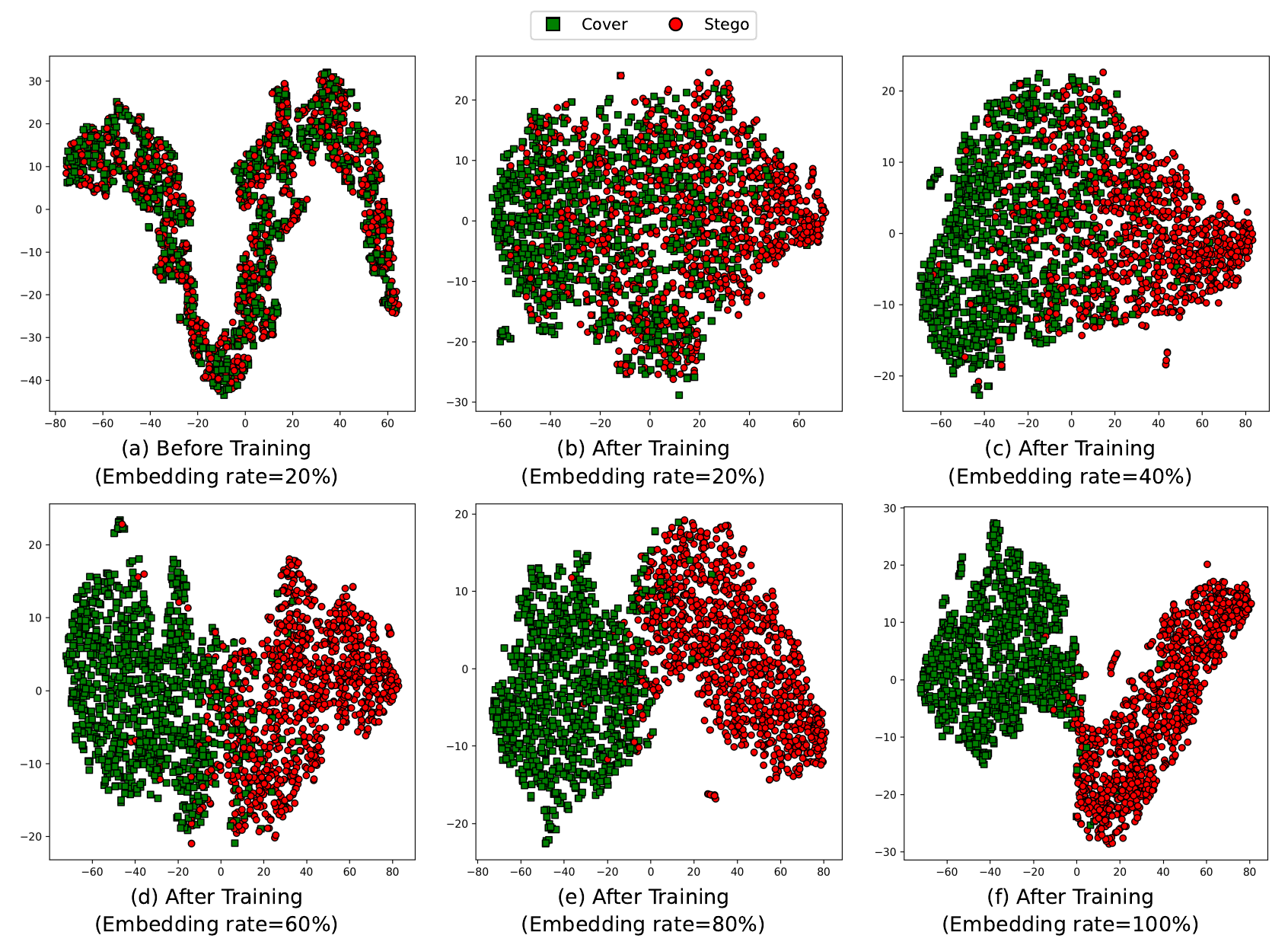}
 \caption{Distribution of graph-level representation vectors before and after training of the proposed model in statistical space across different embedding rates while utilizing a fixed sample length of 0.5 seconds.}
    \label{fig:t-SNE}
\end{figure*}

To further substantiate the discriminative capabilities of our approach, t-distributed stochastic neighbour embedding (t-SNE) was exploited to visualize the graph-level representation vectors (Equation \ref{eq:pooling2}), before and after training, utilizing a set of 2000 graphs selected from the training dataset. Figure \ref{fig:t-SNE} presents snapshots of the distribution of graph-level representation vectors at different embedding rates while utilizing a fixed sample length of 0.5 seconds. {Before training (top-left), the cover and stego data points are heavily overlapped, indicating that they are indistinguishable in their raw feature space.} 
The colour-coded representation illustrates the gradual separation of stego and cover graphs in the vector space of features post-training. The overlap diminishes as the embedding rate increases, reaching almost complete separation at a 100\% embedding rate. {This directly correlates the strength of the steganographic signal with the distance between the clusters in the learned space, visually confirming the model's sensitivity.} These visualizations intuitively showcase the model's efficacy in extracting and analyzing steganographic speech features across various embedding rates.
As a result, the combination of sensitivity to various embedding rates and effectiveness with short samples suggests that our method is versatile and applicable across a wide range of steganographic scenarios, from high-capacity hidden messages to more subtle, security-conscious embeddings. These characteristics position our approach as a powerful tool in the ongoing challenge of audio steganalysis, capable of adapting to different steganographic strategies while maintaining high detection accuracy. These findings provide valuable insights into how the model can be applied in practical VoIP security systems, as discussed in the following section. \textcolor{black}{Moreover, these findings offer valuable insights into the practical applicability of our model in real-world VoIP security systems, which will be further explored in the  section 6.}

\subsection{Comparison with different model variants}

The purpose of this section is to demonstrate the effectiveness of the suggested GraphSAGE-based architecture by comparing it with several of its variants. Six variant architectures, indexed from \#2 to \#7, were considered in the experiment, each with components slightly different from our complete proposed network \#1. The experiment focused on a speech length of 10 seconds and an embedding rate of 20\%

Table \ref{tab:AccVsmodelvaraints} presents the results of the experiment, showing the detection accuracy for each architecture variant. The results highlight the significance of various components in the proposed architecture. It is obvious that the complete model (\(\#1\)) yields the highest level of accuracy, demonstrating the effectiveness of our GraphSAGE for steganalysis. Variant \(\#2\), which removes the first and second max-pooling layers, experiences a significant drop in accuracy, emphasizing the importance of using the hierarchical pooling approach in capturing fine-grained details and high-level patterns. This is particularly beneficial in scenarios with low embedding rates where changes produced by steganography are minimal. In variant \(\#3\), the use of a mean aggregator instead of LSTM highlights the importance of the latter in our approach. The LSTM aggregator captures sequential dependencies and temporal patterns in the input data, which is crucial for steganalysis tasks dealing with speech segments. By leveraging LSTM, the model can effectively learn inter-frame relationships and sequential context,  allowing for more effective detection of subtle changes introduced by QIM steganography. Variant \(\#4\), using max pooling instead of mean pooling, also leads to an accuracy of 85\%, significantly inferior to the complete model. A possible explanation is that max-pooling selects the maximum value from each layer, discarding the information about the overall distribution of features within the layer. In the context of QIM steganalysis, where capturing collective characteristics is crucial for effective detection, this loss of information makes it harder for the model to notice subtle patterns that indicate steganographic content. Mean pooling, on the other hand, averages the features at each layer, providing a more comprehensive representation of the layer's characteristics and proving to be more effective for our specific steganalysis task. Furthermore, variant \(\#5\), where GraphSAGE layers are replaced with GCN, exhibits a notable decrease in accuracy, suggesting that the specific architecture of GraphSAGE contributes significantly to the model's performance. \textcolor{black}{Similarly, when GraphSAGE layers are replaced with graph attention network (GAT) layers (Variant \(\#6\)), while performance is strong (91.63\%), it remains lower than our proposed GraphSAGE model. This indicates that for this specific VoIP steganalysis task, the fixed aggregation approach of GraphSAGE, which comprehensively samples and aggregates features from a fixed neighborhood, provides a more effective representation for QIM-induced subtle changes than the attention-based weighting of GAT layers.}
Finally, in variants \textcolor{black}{\(\#7\)} and \textcolor{black}{\(\#8\)}, the first reduces the number of GraphSAGE layers by one, while the second adds one extra GraphSAGE layer. Both configurations result in a decrease in terms of accuracy, consequently supporting the suitability of employing three layers in our model architecture. Generally, a deeper network has the potential to capture more intricate features within the input data. Variant \textcolor{black}{\(\#8\)}'s performance highlights a crucial observation – the depth of the network does not necessarily correlate with improved model performance. Deeper networks, as seen in variant \textcolor{black}{\(\#8\)}, may face challenges such as overfitting and vanishing gradient problems, emphasizing the importance of striking a balance between network depth and effective steganalysis. The optimal configuration should not only capture relevant features but also mitigate potential issues associated with increased model complexity. 

\textcolor{black}{
This ablation study reaffirms the crucial role each architectural component plays in the overall performance of our model. To further demonstrate the effectiveness of our approach, we next conduct a direct comparison against existing state-of-the-art methods in VoIP steganalysis.}

\begin{table}
    \centering
    \caption{Detection performance under various model configurations.}
    \label{tab:AccVsmodelvaraints}
    \begin{tabular}{@{}lp{4.9cm}l@{}}
        \toprule
         Index& Architecture variant & Accuracy (\%)\\ \midrule
         \#1& The complete proposed model & \textbf{\underline{95.17}} \\
         \#2& Consider only the last mean pooling layer (delete the first and second mean pooling)  & 52.56\\
         \#3& Replace LSTM aggregator with Mean aggregator & 86.70 \\
         \#4& Replace Mean pooling with Max pooling & 85.25\\
         \#5& Replace the GraphSAGE layers with GCN Layers & 68.37\\
         \textcolor{black}{\#6}& \textcolor{black}{Replace the GraphSAGE layers with GAT Layers} & \textcolor{black}{91.63}\\
         \textcolor{black}{\#7}& Use 2 GraphSAGE layers& 94.38\\
         \textcolor{black}{\#8}&  Use 4 GraphSAGE layers& 93.25 \\ \bottomrule
    \end{tabular}
\end{table}

\subsection{Comparison with SOTA methods}

To validate the performance of the proposed model, a comparison was conducted with the following four SOTA methods: IDC \cite{li2012detection}, SS-QCCN \cite{li2017steganalysis}, RNN-SM \cite{lin2018rnn}, and CNN-LSTM \cite{yang2019steganalysis}.
These methods were selected for several reasons. First and foremost, they are specifically tailored for detecting QIM steganography, similar to our approach. Secondly, these methods are established SOTA approaches frequently used as benchmarks in various studies. Lastly, to ensure a fair and equitable comparison, these approaches were reimplemented based on their original papers, which either provided GitHub links for implementation, as with \cite{lin2018rnn}, or were thoroughly documented with sufficient details for independent reimplementation, as provided in \cite{li2012detection, li2017steganalysis, yang2019steganalysis}.

The implementations of these approaches were carried out in Python, adhering to the parameters specified in their respective papers. For IDC and SS-QCCN, given their reliance on SVM with quadratic time complexity, it's impractical to evaluate them using the entire stego and cover segment datasets. Following the experimental settings in \cite{li2017steganalysis}, 4,000 samples were randomly selected for the training set, maintaining a cover-to-stego ratio of 1:1, and 2,000 samples for the testing set, also with a cover-to-stego ratio of 1:1. On the other hand, RNN-SM and CNN-LSTM were evaluated using the same dataset employed in our proposed approach.

\begin{table*}[t]
\centering
\caption{Comparison of detection accuracy with SOTA methods.}
\label{tab:CompSOTA}
\begin{tabular}{@{}llccccc@{}}
\toprule
\multirow{2}{*}{Sample length (s)} & \multirow{2}{*}{Method}                      & \multicolumn{5}{c}{Embedding rate (\%)}                                                        \\ \cmidrule{3-7} 
& & 20 & 40 &60 &80&100 \\ \hline
\multirow{5}{*}{0.5} & IDC \cite{li2012detection}  & 59.15& 73.35& 81.90& 89.40 &94.05  \\  
& SS-QCCN \cite{li2017steganalysis} & 61.25&77.35 & 87.90 &93.15&95.85 \\ 
& RNN-SM \cite{lin2018rnn}  &70.81 &84.39 &\textbf{\underline{93.14}} & 96.09&97.43  \\  
 & CNN-LSTM \cite{yang2019steganalysis}            &\textbf{\underline{71.78}}&\textbf{\underline{85.52}}&92.69&\textbf{\underline{96.76}}& \textbf{\underline{98.29}} \\  
& Our    &70.20&83.35 &92.58 &95.63&98.26 \\ \hline

\multirow{5}{*}{5} & IDC \cite{li2012detection}&68.15 & 84.80& 95.10& 98.35& 99.60    \\ 
& SS-QCCN \cite{li2017steganalysis}&71.30 &96.15& 99.40&99.90&\underline{\textbf{100}}    \\  
& RNN-SM \cite{lin2018rnn}  & 67.91& 88.76&96.59   &98.67&99.60 \\ 
& CNN-LSTM \cite{yang2019steganalysis}            &89.98 &96.80 & 99.66 &99.81&99.87  \\  
& Our    & \textbf{\underline{90.76}}& \textbf{\underline{98.88}}&\textbf{\underline{99.89}}&\textbf{\underline{99.97}}&99.98 \\ \hline

\multirow{5}{*}{10} & IDC \cite{li2012detection}&64.50 & 85.60&95.95& 99.55& 99.70     \\ 
& SS-QCCN \cite{li2017steganalysis}           & 72.15&98.75& 99.80&100&100    \\ 
& RNN-SM \cite{lin2018rnn}  & 62.41& 92.74& 98.32  &99.65&99.93 \\ 
& CNN-LSTM \cite{yang2019steganalysis}&92.38 &99.65 &  99.74 &99.93&99.96 \\ 
& Our    &\textbf{\underline{95.17}} & \textbf{\underline{99.68}}&\textbf{\underline{99.94}}&\textbf{\underline{100}}&\textbf{\underline{100}} \\ \bottomrule
\end{tabular}

\end{table*}

Table~\ref{tab:CompSOTA} presents a detailed comparison of detection accuracy between the proposed model and SOTA approaches across various speech segment lengths and embedding rate settings. 
The results reveal that our approach and CNN-LSTM exhibit superior accuracy. Our model demonstrates competitive performance in 0.5-second and 5-second speech segments, compared to RNN-SM and SS-QCCN, respectively. In comparison to IDC and SS-QCCN, our approach significantly outperforms these methods, notably in cases where embedding rates are low and segments are short. When compared to RNN-SM, our approach demonstrates a notable disparity, especially at low embedding rates  (20\% and 40\%). However, in the context of 0.5-second speech length, RNN-SM performs on par or slightly better than our model. Overall, across all scenarios where our model does not achieve the highest accuracy, its performance remains  close to the best results.

\begin{figure*}
    \centering
    \includegraphics[scale=0.40]{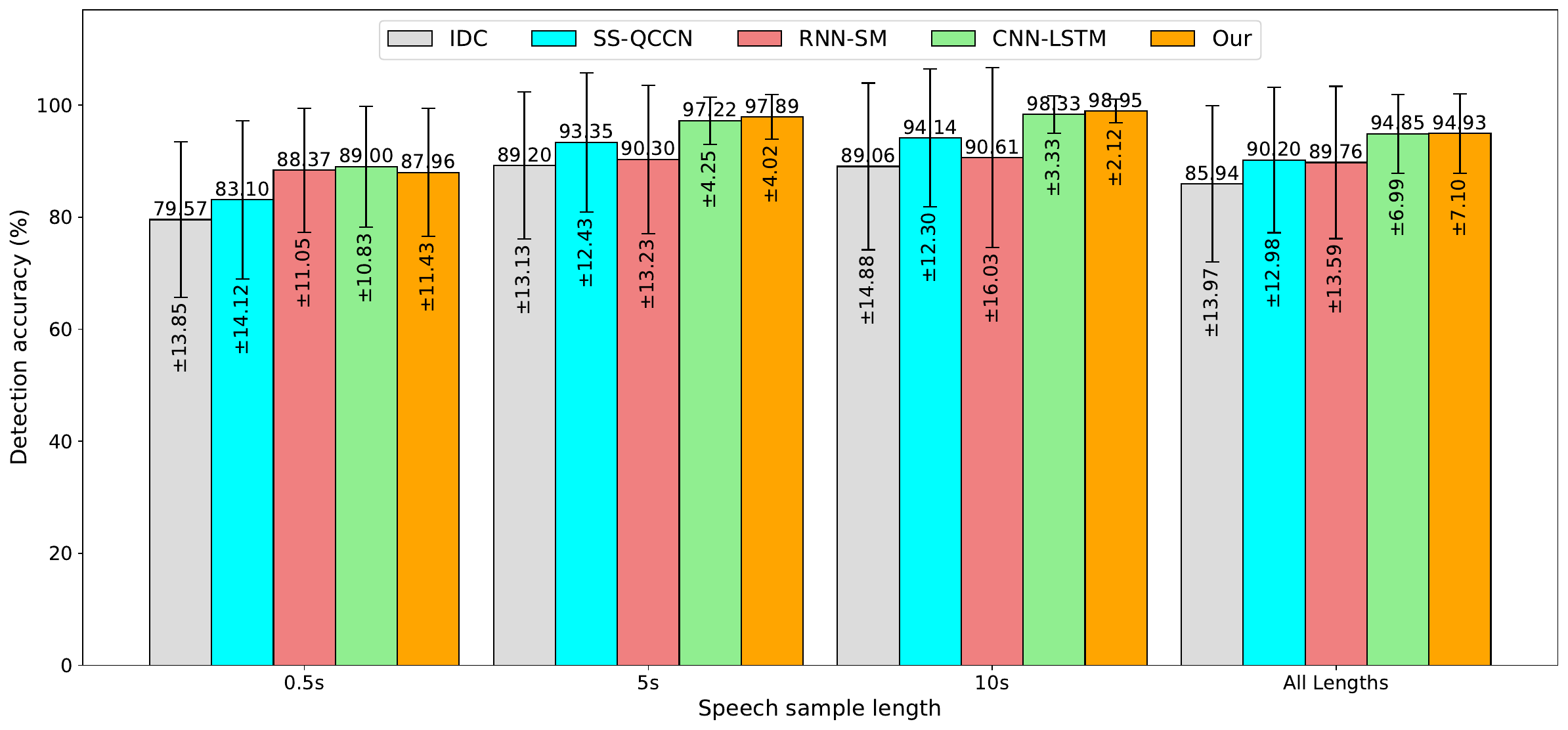}
    \caption{Comparison of mean detection accuracy with standard deviation for different methods across various sample lengths.}
    \label{fig:statAccu}
\end{figure*}

Additionally, when the sample duration is 5 seconds and 10 seconds, and the embedding rate surpasses 80\%, the detection accuracy of almost all models stabilizes around 99\%. The primary challenge arises in scenarios with short speech fragments and/or low embedding rates, where our approach still demonstrates satisfactory performance. In terms of statistics, as presented in Figure \ref{fig:statAccu}, our method demonstrates competitive and, in certain scenarios, superior performance in terms of mean detection accuracy and standard deviation compared to other SOTA methods. At shorter sample lengths (0.5 seconds), while our mean accuracy slightly trails behind the CNN-LSTM method, it maintains a lower standard deviation, indicating more consistent performance across different embedding rates. As the sample length increases to 5 and 10 seconds, our method outperforms all other techniques, achieving the highest mean accuracy and standard deviation. Furthermore, when considering the overall performance across all sample lengths, our method exhibits the highest mean accuracy and the most consistent performance.

To further validate the approach, a detection comparative analysis was conducted with \textcolor{black}{four} recently proposed methods: \textcolor{black}{ E-SWAN \cite{wang2025swan},  Bi-LSTM-3DCNN \cite{li2025compressed}}, TENet \cite{zhang2023tenet}, and FedSpey \cite{tian2023fedspy}. These methods use the same dataset and QIM steganography technique as our work. Due to the complexity of the techniques and limited available information, these methods were not re-implemented. Consequently, the comparison is based on results presented in the corresponding published papers and is limited to scenarios common to our study.

For the TENet approach, which is based on Transformer architecture, the nearest comparable scenario to our study involves embedding rates of 50\% with sample lengths of 1, 5, and 10 seconds. This closely aligns with our scenario of 60\% embedding rates and identical sample lengths. TENet reports accuracies of 96\%, 99\%, and 99\% for these respective durations. In comparison, our approach achieves accuracies of 95.83\%, 99.89\%, and 99.94\% for the same sample lengths.

FedSpy applies federated learning to several SOTA steganalysis approaches, specifically RNN-SM. Their results focus on 1-second sample lengths with varying embedding rates. For a 40\% embedding rate, FedSpy achieves an accuracy of about 95\%, while our approach attains 90.98\%. At a 100\% embedding rate, both approaches reach near 100\% accuracy.  

\color{black}
E-SWAN, which relies on LSTM and convolutional modules, was evaluated using only 10-second audio samples across varying embedding rates. At a low embedding rate of 20\%, it achieved an accuracy of 94.65\%, while our approach reached 95.17\%.

The Bi-LSTM-3DCNN method was tested with 10-second samples at different embedding rates and with 100\% embedding at varying sample lengths. In the challenging scenarios---10 seconds at 20\% embedding and 1 second at 100\% embedding---it achieved accuracies of 77.99\% and 80.61\%, respectively. In contrast, our approach significantly outperformed it, achieving 95.17\% and 99.47\% in the same settings.

\color{black}
These findings underscore the robustness and effectiveness of our proposed GraphSAGE-based model across diverse scenarios, establishing it as an alternative and advanced solution in QIM steganalysis.

\subsection{Time efficiency \textcolor{black}{and compuational complexity}}
\begin{figure}[t!]
    \centering
    \includegraphics[scale=0.6]{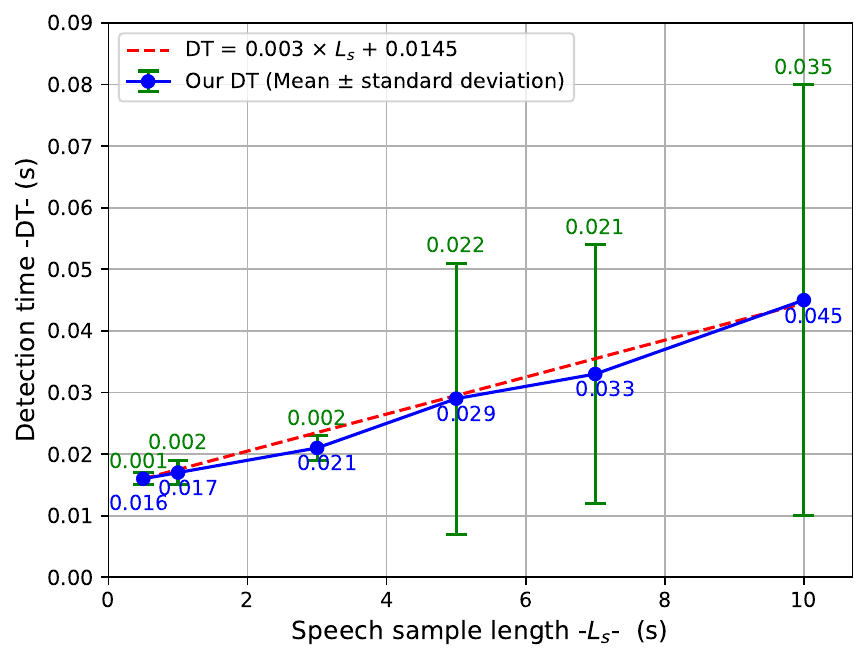}
    \caption{DT of the proposed GNN approach across various sample lengths.}
    \label{fig:DetectionTime}
\end{figure}

Time efficiency is a crucial consideration when assessing the practical applicability of a model, particularly for online applications. To evaluate the time efficiency of our method, the average DT across various sample lengths was computed. The detection process is depicted in Figure \ref{fig:TrainTestProcess} (right-hand process). Furthermore, a comparative analysis was conducted with the four previously mentioned approaches to assess the efficiency of our method in relation to existing steganalysis models. The detection process considered in our comparison encompasses the phases starting with a compressed VoIP sample and concluding with the decision on whether a secret message is embedded. \textcolor{black}{It is important to emphasize that all compared models, including ours, are trained offline. During the detection phase, the pre-trained models are loaded into memory and used for inference, which reflects a realistic deployment scenario. This ensures a fair comparison focused solely on runtime detection performance.}
These experiments were implemented on the Kaggle platform, utilizing only CPU without any GPU acceleration. To ensure a fair comparison, the mean and standard deviation were computed over 200 detection tests for each sample length \(L_s\). Figure \ref{fig:DetectionTime} presents the DT of our approach across different sample lengths. Notably, the mean DT demonstrates a nearly linear increase with respect to the sample length, as approximately modelled by Equation \ref{dt}, expressed as follows:

\begin{equation}
    \text{DT} = 0.003  \times L_s + 0.0145
    \label{dt}
\end{equation}

The standard deviation, visualized by the arrows, highlights the variability in DTs, particularly for longer samples. 

It is evident that our GNN model maintains a relatively consistent mean detection time, even when processing moderately longer speech segments. For instance, detecting steganography in a 10-second sample takes approximately 0.045 seconds, representing only 0.45\% of the sample length. While detecting steganography in short segments poses a challenge, our approach ensures fast detection in such scenarios. Detecting steganography in 0.5-second samples takes about 0.016 seconds, which represents approximately 3.2\% of the sample length. These performances are attributed to several factors. Firstly, the adopted GraphSAGE architecture is relatively simple, leading to a reduced computational load. This simplicity contributes to faster training and DTs. Secondly, the GraphSAGE layer uses graph-based operations for information aggregation. The method of aggregating information from neighbouring nodes in a graph can be computationally efficient, especially for tasks that involve capturing dependencies and relationships in a graph structure. Finally, the simplicity of the graph construction approach reduces the computational complexity and resource requirements both in the training and detection phases. These results underscore the efficiency of our suggested GNN-based steganalysis method, demonstrating its suitability for deployment in online steganalysis tasks.

Table \ref{tab:DetectTime} provides insights into the time efficiency of our approach compared to SOTA methods in three sample length scenarios (0.5 seconds, 5 seconds, and 10 seconds). The detection time is reported as the mean value [\(\pm\) standard deviation]. Our approach consistently demonstrates superior time efficiency, outperforming the SOTA methods in terms of speed across all sample lengths. The IDC approach presents closer DTs to our approach, which can be explained by the simple architecture adopted in IDC, focusing only on inter-frame correlation. SS-QCCN, on the other hand, yields the least favourable results, primarily due to its method of extracting intra- and inter-frame correlations and its utilization of the PCA method to reduce the dimensionality of a 131,072-dimensional feature vector to 300. This process incurs a significant amount of time, contributing to the observed extended DT in this method. For CNN-LSTM and RNN-SM, the complicated architecture, particularly for CNN-LSTM, which involves sequential and convolutional operations, might be more computationally intensive, especially when dealing with longer sequences, leading to slower DTs compared to ours. These results demonstrate the exceptional efficiency of our proposed model, enabling near real-time identification of hidden data in VoIP voice transmissions. This efficiency has significant practical implications, allowing for continuous monitoring of communications and immediate detection of steganographic content. The model's scalability across various sample lengths and its superior speed compared to SOTA methods indicate excellent adaptability and resource efficiency. This performance could significantly enhance operational efficiency in security applications and allow for potential seamless integration into existing VoIP infrastructure.

\begin{table*}[t!]
\centering
\caption{DT (s) of our approach compared to SOTA method. Results are reported as Mean [\(\pm\) standard deviation]}
\label{tab:DetectTime}
\begin{tabular}{@{}lccc@{}}
\toprule
\multirow{2}{*}{Method} & \multicolumn{3}{c}{Sample lenght (s)}   \\ \cmidrule{2-4}
 & 0.5    & 5   & 10     \\\midrule
IDC \cite{li2012detection}&    0.019 [\(\pm\) 0.004]         & 0.037 [\(\pm\) 0.003]          &  0.049 [\(\pm\) 0.002]          \\
SS-QCCN \cite{li2017steganalysis}&  0.099 [\(\pm\) 0.056]            &   0.134 [\(\pm\) 0.070]       &       0.146 [\(\pm\) 0.053]     \\
RNN-SM \cite{lin2018rnn}& 0.032 [\(\pm\) 0.057]              &    0.081 [\(\pm\) 0.083]      & 0.127 [\(\pm\) 0.075]           \\
CNN-LSTM \cite{yang2019steganalysis}&  0.075 [\(\pm\) 0.069]            &      0.119 [\(\pm\) 0.068]    &  0.152 [\(\pm\) 0.083]          \\
Our                     &  \textbf{\underline{0.016  [\(\pm\) 0.001]}}  &  \textbf{ \underline{0.029 [\(\pm\) 0.022]}}       & \textbf{\underline{0.045 [\(\pm\) 0.035]}} \\ \bottomrule
\end{tabular}
\end{table*}

\color{black}
While our reported DT in Table \ref{tab:DetectTime} and Figure \ref{fig:DetectionTime} focuses on single-sample inference—a common practice in related steganalysis works \cite{yang2019hierarchical,yang2022real,li2022general,lin2018rnn,yang2019steganalysis,li2012detection,li2017steganalysis,wu2020steganalysis,yang2018steganalysis,yang2019fast,hu2021detection,wang2021fast,li2021detection,wang2022steganalysis,li2023sanet,wang2025swan,li2025compressed,yang2020fcem,qiu2022steganalysis,zhang2023tenet,tian2023fedspy,zhang2024spm} that assesses minimal latency for individual streams—the architecture of our proposed GraphSAGE-based model is inherently well-suited for batch processing, a critical aspect for high-throughput real-world deployments. The core operations within our GraphSAGE layers are highly parallelizable, meaning multiple sliding windows (each representing a speech segment) can be grouped into batches and processed simultaneously. This capability, efficiently managed by modern deep learning frameworks on GPUs or multi-core CPUs, allows for a significant reduction in the effective detection time per sample when handling a high volume of traffic. Thus, the low single-sample DT provides a strong foundation for our system's scalability and high throughput in large-scale, real-time VoIP environments.

To further validate our system's efficiency, we also conducted an analysis of the computational complexity for all evaluated models, quantifying them by their number of parameters and floating point operations (FLOPs), as presented in Table \ref{tab:flops}. Although our model does not have the lowest FLOPs, it maintains a strong balance between architectural simplicity and performance. It uses fewer parameters than IDC, SS-QCCN and CNN-LSTM while achieving a significantly faster DT. Several key architectural and implementation factors explain this outcome.  While FLOPs quantify raw arithmetic operations, the actual execution speed is heavily influenced by factors like memory access patterns, instruction pipelining, and how efficiently operations can be vectorized or parallelized by the underlying hardware and software libraries (e.g., BLAS for matrix multiplications). IDC's very low FLOPs (1.53 MFLOPs) stem from its statistical feature extraction, which involves computations like histograms and Markov chain probabilities. These operations, while theoretically low in FLOPs, can involve frequent, less cache-friendly memory accesses or branch predictions that are less optimized for modern CPUs compared to highly parallelizable matrix multiplications typical in GNN layers. GraphSAGE, especially with the simple directed acyclic graph structure we employ, might be particularly efficient for single-sample inference on CPUs. The aggregation and update steps, while involving more FLOPs than a direct statistical lookup, may benefit from more optimized library calls that process data blocks efficiently. RNN-SM, while having fewer parameters and FLOPs, might be subject to the inherent sequential bottleneck of RNNs, making it slower for CPU inference on single, short samples compared to our parallelizable graph operations. 

The efficiency of our model lies not just in theoretical FLOPs, but in the overall streamlined pipeline, the nature of its core computational blocks which are well-suited for modern CPU architectures, and its ability to capture complex relational features without relying on computationally expensive preliminary feature engineering common in some other methods.
\color{black}

\begin{table}
\color{black}
    \centering
    \caption{\textcolor{black}{Comparison of model parameters and FLOPs for all methods}}
    \label{tab:flops}
    \begin{tabular}{lll}
        \toprule
         \textbf{Model}& \textbf{Parameters} & \textbf{FLOPS}\\ \midrule
         IDC \cite{li2012detection}& 735K & 1.53M  \\
         SS-QCCN \cite{li2017steganalysis}& 5.2M  &11.63M \\
         RNN-SM \cite{lin2018rnn}& 33K &6.1M  \\
         CNN-LSTM \cite{yang2019steganalysis}& 258K & 46.2M\\
         Our & 83K& 22.8M \\ \bottomrule
    \end{tabular}
    \color{black}
\end{table}

\section{Real-world applications, limitations, and improvements }\label{section6}

In real-world applications, the proposed GraphSAGE-based steganalysis approach offers substantial practical benefits for enhancing the security of VoIP communication systems and safeguarding against covert communication channels. Its key strengths—high accuracy and efficiency—make it particularly well-suited for deployment in various practical scenarios.

\setlength{\leftmargini}{10pt}
\begin{itemize}
    \item \textbf{Cybersecurity and network monitoring:} Internet service providers (ISPs) and network administrators could deploy this system to detect hidden communications in VoIP traffic, potentially uncovering malicious activities or data exfiltration attempts. Corporate networks could use it to ensure compliance with data protection policies and prevent unauthorized data transfers.
    
    \item \textbf{Law enforcement:} Intelligence agencies could utilize this technology to identify covert communication channels used by criminal organizations or terrorist groups. Digital forensics teams could apply this method to analyze seized communication devices for hidden messages.
   
    \item \textbf{Digital rights management:} Content distribution platforms could implement this approach to detect unauthorized watermarking or copyright infringement in audio streams.
    
    \item \textbf{E-learning and online examination integrity:} Educational institutions could use this technology to ensure the authenticity of voice-based online assessments and prevent cheating through hidden audio cues.
    
    \item \textbf{IoT security:} As voice-controlled IoT devices become more prevalent, this approach could be used to detect potential security breaches or unauthorized access attempts via steganographic commands.
\end{itemize}
\vspace{0.3cm}

Although our approach demonstrates strong performance, it faces some limitations and challenges. Firstly, the model struggles with very short sample lengths (less than 0.5 seconds) and extremely low embedding rates (below 20\%). This difficulty arises because there is limited information available in these cases, making it challenging to extract enough features for accurate detection.
One possible solution to overcome this limitation is to integrate GAT \cite{velickovic2017graph} into the model. GATs are a type of GNN that utilizes attention mechanisms to focus selectively on the most relevant connections and information within the graph. By assigning higher weights to connections that reveal significant variations in QIM sequences, GATs "could be" more effectively capture subtle changes in short samples and low embedding rates. 

Secondly, the current model is specifically designed for QIM-based steganography \textcolor{black}{in G.729 compressed speech}, and its effectiveness in detecting other steganography methods used in VoIP streams, such as those based on ACB, FCB \textcolor{black}{ and LSB, or in different codecs (e.g., G.711, G.722)}, is limited. To address this limitation and enhance its versatility, a multi-graph construction approach combined with a fusion network can be explored. 
This involves constructing separate graphs \textcolor{black}{for different parameter types (QIS, ACB, FCB, etc.) or even for representations derived from different codecs,} and then capturing the unique variations introduced by each steganography method \textcolor{black}{or codec}. A fusion network, potentially implemented using LSTM or attention mechanisms, would then aggregate and fuse features extracted from these individual graphs, learning to dynamically weigh their contributions. This combined approach could enable the model to generalize to different steganography techniques \textcolor{black}{and codecs} and enhance its ability to detect a broader range of hidden messages within VoIP streams.


\color{black}
Finaly, although the dataset is diverse in terms of speakers and language (English and Chinese), it is based on pre-recorded data and a specific codec/steganography pair. Real-world VoIP traffic can exhibit  variability in network conditions, background noise, codec implementations, and potentially unknown or adaptive steganography techniques. Evaluating the model's performance on live or more diverse traffic is necessary.
\color{black}

\color{black}
\section{Ethical considerations}
The primary objective of this research is to enhance cybersecurity by providing a robust and efficient method for detecting covert communications hidden within VoIP streams. This capability is crucial for defensive purposes, including protecting national security, preventing intellectual property exfiltration, and aiding law enforcement in countering illicit activities. However, we acknowledge that, like many cyberscurity technologies, steganalysis tools could potentially be misused for unauthorized surveillance or infringement on individual privacy.

It is important to emphasize that our method is designed solely to detect the presence of steganography, not to extract, interpret, or monitor the content of any communication. Its responsible deployment must be strictly governed by strong legal frameworks and transparent ethical guidelines to ensure that it is used only for legitimate purposes and in full compliance with privacy laws and human rights. Our contribution aims to empower defenders in the ongoing cyber security landscape, fostering a more secure digital environment where covert malicious activities can be identified, thus supporting a balance between security needs and individual privacy. We advocate for the ethical application of this technology in accordance with all relevant regulations and principles.
\color{black}

\section{Conclusion}
\label{section7}
In this article, a novel steganalysis algorithm for VoIP streams based on the GraphSAGE architecture was introduced. 
The proposed GNN method showcases efficient performance in detecting QIM steganography in VoIP signals. Notably, when compared to existing SOTA algorithms, it demonstrates a superior compromise between detection accuracy and efficiency.  \textcolor{black}{Achieving detection accuracy exceeding 98\% for 0.5-second samples, and 95.17\% under 20\% embedding rate scenarios— representing an improvement of 2.8\% over the best-performing SOTA approaches. Additionally, it maintains high computational efficiency, with an average detection time as low as 0.016 seconds (a 0.003-second improvement), which corresponds to less than 3\% of the sample length, making it well-suited for online detection systems.}
Additionally, this work contributes to the field by introducing the use of GNNs, specifically GraphSAGE, in VoIP steganalysis, showcasing its practicality and effectiveness.

Looking ahead, future research will concentrate on refining the model to address its limitations as detailed in section \ref{section6}. This includes enhancing its ability to detect very short sample lengths with very low embedding rates, handling various steganography methods, and accurately predicting embedding rates. Additionally, an important direction for future work is to extend the model to identify the specific positions of embedded bits within the VoIP streams. This challenging scenario may involve developing more granular feature extraction techniques that focus on bit-level analysis. \textcolor{black}{Furthermore, investigating the robustness of GNN-based steganalysis against evasion and adversarial attacks remains an essential research avenue. Incorporating adversarial training strategies could help improve the model’s resilience to sophisticated obfuscation attempts. Finally, extending the proposed GNN-based framework to multi-modal steganalysis — for example, combining VoIP with video streams — could broaden its applicability and impact, paving the way for more comprehensive cross-domain steganalysis systems.}



\section*{Acknowledgements}
The open access (OA) fee for this paper was funded by the University of Liverpool.

\section*{Declaration of Competing Interest}
No competing financial interests or personal affiliations are associated with the authors that could have impacted the content of this paper.

\section*{Data Availability}

The data that support the ﬁndings of this study are available upon request from the corresponding author.

\balance
\bibliographystyle{elsarticle-num} 
\bibliography{Bibliography}

\end{document}